\documentclass[a4paper,11pt]{article}
\usepackage[utf8]{inputenc}
\usepackage[longnamesfirst]{natbib}
\usepackage[top=35truemm,bottom=35truemm,left=27truemm,right=27truemm]{geometry}

\usepackage{amsthm, amssymb, amsmath, appendix, url}
\usepackage{physics}
\usepackage{mathtools}
\mathtoolsset{showonlyrefs=true} 

\theoremstyle{definition}
\newtheorem{theorem}{Theorem}
\newtheorem{lem}{Lemma}

\newtheorem{proposition}{Proposition}

\newtheorem{remark}{Remark}

\newtheorem{example}{Example}

\newcommand{\argmax}{\mathop{\rm arg ~ max}\limits}

\usepackage{tikz}
\usetikzlibrary{intersections, calc, arrows.meta}

\makeatletter

\@addtoreset{figure}{section}
\@addtoreset{table}{section}
\makeatother 

\title{\huge{Social Choice Rules with Responsibility for Individual Skills}}
\author{Kensei Nakamura\thanks{Graduate School of Economics, Hitotsubashi University, Kunitachi, Tokyo 186-8601, Japan. E-mail: kensei.nakamura.econ@gmail.com (ORCID: 0009-0008-4549-4215)}}
\date{This version : \today}

\begin{document}

\maketitle

\vspace{5mm}

\begin{abstract}
This paper examines normatively acceptable criteria for evaluating social states when individuals are responsible for their skills or productivity and these factors should be accounted for. We consider social choice rules over sets of feasible utility vectors à la Nash's (\citeyear{nash1950bargaining}) bargaining problem. First, we identify necessary and sufficient conditions for choice rules to be rationalized by welfare orderings or functions over ability-normalized utility vectors. These general results provide a foundation for exploring novel choice rules with the normalization and providing their axiomatic foundations. By adding natural axioms, we propose and axiomatize a new class of choice rules, which can be viewed as combinations of three key principles: distribution according to individuals' abilities, utilitarianism, and  egalitarianism. Furthermore, we show that at the axiomatic level, this class of choice rules is closely related to the classical bargaining solution introduced by \citet{kalai1975other}. 
\end{abstract}

\newpage
%%%%%%%%%%%%%%%%%%%%%%%%%%%%%%%%%%%%%%%%%%%%%%%%%%%
\section{Introduction}
\label{sec_intro}
%%%%%%%%%%%%%%%%%%%%%%%%%%%%%%%%%%%%%%%%%%%%%%%%%%%

If individuals are responsible for their skills or productivity, these factors should be taken into account when evaluating social states or outcomes. 
Naturally, whether this standpoint is reasonable or not is controversial, and the answer depends on the specific situations under consideration. 
However, if we focus on situations where differences in talent are negligible and equality of opportunity has been achieved, then abilities can be viewed as a reflection of the effort individuals have exerted and the extent to which they have been industrious or lazy. 
In these cases, it is natural, or perhaps necessary, to consider these factors when evaluating social states.

Suppose there is a fixed amount of capital and two individuals, Ann and Bob, divide and use it to produce their own consumption goods. 
Let us assume that Ann’s productivity is significantly higher than Bob’s. 
If the difference in skills arises from factors such as innate talents or their parents’ incomes, then the capital should be allocated to compensate for the disadvantage, meaning Bob should receive more capital (\citealp{fleurbaey1995equality}). 
On the other hand, if the difference in their talents is very small and they have been given the same opportunities to improve their skills---for instance, education---then we can infer that Bob has spent less effort than Ann, and conclude that Bob should be held responsible for his past (\citealp{arneson1989equality}). 
In this example, ex-post equal treatment is not appropriate; in other words, it would be undesirable to allocate additional capital to Bob to compensate for his lower skills. Instead, it would be preferable for the chosen distribution to reflect the difference in their abilities, giving Ann more capital than she would receive if she were not considered responsible for her abilities.
This example motivates us to explore desirable ability-dependent criteria from a normative perspective.

To explore desirable criteria when individuals are responsible for their abilities, this paper axiomatically examines choice rules depending on sets of feasible utility (or payoff) vectors à la Nash's (\citeyear{nash1950bargaining}) bargaining problem. 
For simplicity, we interpret the maximum utility levels they can achieve as their abilities and characterize several choice rules that account for individual abilities. 
(In the example above, sets of feasible payoff vectors are derived from the amount of consumption goods that they can produce from each capital allocation.
Each individual's ability is the level of consumption goods they could achieve if they had the power to dictate.)

First, we axiomatize general classes of choice rules written as the maximizers of ability-normalized welfare orderings or functions  with several basic properties. 
These characterization results are obtained from two axioms related to how the social planner takes the distribution of abilities into account, together with standard axioms such as efficiency, anonymity, and continuity. 
One postulates the invariance of choice under the condition that all individuals have equal abilities both before and after changes in feasible sets. 
This axiom can be seen as a restricted version of Arrow's choice axiom (\citealp{arrow1959rational}) or Nash's independence of irrelevant alternatives (IIA). 
The other is the scale-invariance property, which requires that if the utility levels some individuals can achieve are changed, then their utility levels in the chosen outcomes should change proportionally.  
Since these results offer very general characterizations, they have broad applications.

By considering natural and acceptable axioms in addition to ones in the first general results, we propose and axiomatize a new class of choice rules that can be seen as combinations of three key principles: \textbf{\textit{distribution according to individuals' abilities, utilitarianism, and  egalitarianism}}.
In general, benevolent social planners place greater emphasis on disadvantaged individuals, and the degree of this can be represented as a weight over individuals. 
Our new choice rules can be interpreted as ones of an inequality-averse planner  with a symmetric set $\mathcal{W}$ of weights in her mind. 
The planner evaluates each utility vector as follows:  After normalizing the individuals' utility levels by their abilities, she computes the weighted sum of the normalized utility levels with respect to each plausible weight $w\in \mathcal{W}$ (in a utilitarian way), and she chooses the smallest weighted sum as the utility vector evaluation. 
Since the final step gives higher weights to disadvantaged individuals, it captures an egalitarian attitude of the social planner.\footnote{The weight used to evaluate each utility vector is determined by the normalized utility levels. Hence, this rule sometimes assigns higher weights to individuals who attain high utility levels when their maximum utility levels are also high enough. Even in such a situation, we can interpret this operation as a fair one in a sense since individuals with low normalized utility levels can be deemed as unrewarded relative to their effort.}
We refer to these as  the \textbf{\textit{relative fair choice rules}}.

The weight set represents the extent to which the social planner is inequality-averse. 
When the weight set equals the entire set, the corresponding relative fair choice rule evaluates outcomes by the maximin rule (\citealp{Rawls1971-JOHATO-12}) after normalizing utility levels by individuals' abilities. 
We refer to this as the \textit{relative maximin choice rule}. 
This choice rule closely resembles the Kalai-Smorodinsky (KS) choice rule (\citealp{kalai1975other}), which has been widely examined in axiomatic bargaining theory: Both rules can be viewed as maximizing the same objective function and select the same outcomes in convex problems. 
The only difference is whether or not ties are broken: The KS choice rule chooses only outcomes proportional to their abilities.

On the other hand, when the weight set is a singleton,  this choice rule becomes the utilitarian rule (\citealp{Bentham1780-BENITT}) with the  normalization. 
It chooses the maximizers of the sum of utility levels normalized by each individual's ability.  
This concept  is rooted in \textit{relative utilitarianism} in social choice theory (e.g., \citealp{dhillon1999relative,karni1998impartiality,segal2000let,sprumont2019relative}), and we refer to this as the \textit{relative utilitarian choice rule}. 
% This has been examined in axiomatic bargaining theory as well (e.g., \citealp{baris2018timing,peitler2023rational,pivato2009twofold}). 

We provide new axiomatic foundations for these two extreme cases. These characterizations can be derived simply by combining our general characterization with existing results in social choice theory. This highlights the significance of the general results presented in the first part. 
Additionally, we discuss other notable choice rules within the class of relative fair choice rules.

Furthermore, we show that the relative fair choice rules are closely related to the KS choice rule at the axiomatic level as well. 
The main differences between these choice rules lie in the axioms of efficiency and symmetry. The KS choice rule satisfies all the axioms in our characterization of the relative fair choice rules except for an efficiency axiom, and if we weaken it and strengthen the axiom of symmetry, then the KS choice rule can be obtained. 
Moreover, we prove that in the two-person case, these two choice rule concepts can be jointly characterized by relaxing the efficiency axiom in our characterization of the relative fair choice rules.

This paper is organized as follows: Section \ref{sec_brgainingProblem} introduces the setup, and Section \ref{sec_axioms} identifies the necessary and sufficient conditions for a choice rule to be written as the maximizers of an ability-normalized welfare ordering or function with several basic properties. 
Section \ref{sec_modifiedKS} provides the characterization of the relative fair choice rules.
Section \ref{sec_specialcases} discusses special cases of the relative fair choice rules and axiomatizes the two extreme cases. 
Section \ref{sec_comparison} examines the axiomatic relationship between the relative fair choice rules and the KS choice rule. 
Section \ref{sec_literature} reviews the related literature, and Section \ref{sec_discussion} provides concluding remarks. All proofs are in Appendix. 

%%%%%%%%%%%%%%%%%%%%%%%%%%%%%%%%%%%%%%%%%%%%%%%%%%%
\section{The Social Choice Problem}
\label{sec_brgainingProblem}
%%%%%%%%%%%%%%%%%%%%%%%%%%%%%%%%%%%%%%%%%%%%%%%%%%%

Let $N = \{1,2, \cdots, n \} $ be the set of individuals with $n\geq 2$. 
Let $X \subset \mathbb{R}_+^n$ be a set of feasible utility vectors.\footnote{
Let $\mathbb{R}$ (resp. $\mathbb{R}_{+}, ~ \mathbb{R}_{++}$) denote the set of real numbers (resp. nonnegative numbers, positive numbers). Let $\mathbb{R}^n$ (resp. $\mathbb{R}^n_{+}, ~ \mathbb{R}^n_{++}$) denote the $n$-fold Cartesian product of $\mathbb{R}$ (resp. $\mathbb{R}_{+}, ~ \mathbb{R}_{++} $). Let $\mathbb{N}$ be the set of natural numbers.
}
We assume that utility gains and levels are interpersonally comparable.
The utility vector of the status quo is  $(0,0, \cdots 0)$, denoted by $\mathbf{0}$. 
A \textit{(social choice) problem} is a set $X \subset \mathbb{R}^n_{+}$ that satisfies the following conditions:
\begin{itemize}
    \item The set $X$ is nonempty, compact, and comprehensive (i.e., for all $x, y \in \mathbb{R}^n_+$, if $x \in X$ and $y\leq x $, then $y\in  X$).\footnote{We write $a \gg b$ if $a_i > b_i$ for all $i\in N$, and $a \geq b$ if $a_i \geq b_i$ for all $i\in N$. We define $\ll$ and $\leq$ in the same way. }  
    \item For all $i\in N$, there exists $x\in X$ such that $x_i > 0$. 
\end{itemize}
% Note that we do not impose  the convexity on problems.\footnote{
% The convexity assumption is standard but justified by the randomization of outcomes.
% To examine situations where the randomization is not valid or prohibited, we include non-convex problems into the domain. Many papers also do not assume the convexity
% (e.g., \citealp{conley1991bargaining,herrero1989nash,mariotti2000maximal,ok1999revealed,xu2006alternative,zhou1997nash}).
% For a comprehensive survey of non-convex problems, see Section 4 of \citet{Thomson2022RED} and \citet{xu2020nonconvex}.  
% } 
The set of problems is denoted by $\mathcal{P}$.

A \textit{choice rule} $F$ assigns a nonempty subset $F(X)$ of $X$ for each problem $X\in \mathcal{P}$. 
For all $X\in \mathcal{P}$, define $b(X) \in \mathbb{R}^n_{++}$ as $b_i (X) = \max_{x\in X} x_i$ for each $i\in N$.  
Given $X\in \mathcal{P}$, we interpret $b_i (X)$ as individual $i$'s ability. 
We restrict our attention to situations where differences in talents are negligible and equality of opportunity has been achieved.
Under these conditions, abilities can be seen as depending on the effort individuals exert and the extent to which they are industrious or lazy, so individuals are responsible for their abilities. 
This paper examines choice rules that take $b(X)$ into account to reflect differences in abilities when evaluating feasible vectors.

Let $\mathbf{1}$ denote $(1 , 1, \cdots, 1) \in \mathbb{R}^n$. 
A function $\pi: N\rightarrow N$ is a \textit{permutation} if it is a one-to-one function. Let $\Pi$ be the set of permutations. 
For all $x\in \mathbb{R}^n$, let $x^\pi = (x_{\pi (1)}, x_{\pi (2)}, \cdots, x_{\pi (n)})$. 
We say that a set $A\subset \mathbb{R}^n$ is \textit{symmetric} if  for all  $\pi \in \Pi$, $A =\{ x^\pi \mid x\in A\}$. 
For all $x\in \mathbb{R}^n$, $(x_{(1)}, x_{(2)}, \cdots, x_{(n)})$ denotes a rearrangement of $x$ such that $x_{(1)}\leq  x_{(2)}\leq  \cdots\leq x_{(n)}$, with the ties being broken arbitrarily. 
% We say that a set $A\subset \mathbb{R}^n$ is \textit{symmetric} if  for all permutations $\pi$, $A =A^\pi$. 
For all $a,x  \in \mathbb{R}^n$, let $ax = (a_1 x_1, a_2 x_2, \cdots a_n x_n)$. For all $a\in \mathbb{R}^n$ and $X\subset \mathbb{R}^n$, let $aX = \{ ax \mid x\in X \}$.
Additionally, for all $X \subset \mathbb{R}^n$ and $\alpha \in \mathbb{R}$, let $X + \alpha \mathbf{1} = \{  x + \alpha \mathbf{1} \mid x \in X   \}$. 

\begin{figure}
    \centering
    \includegraphics[width=0.8\linewidth]{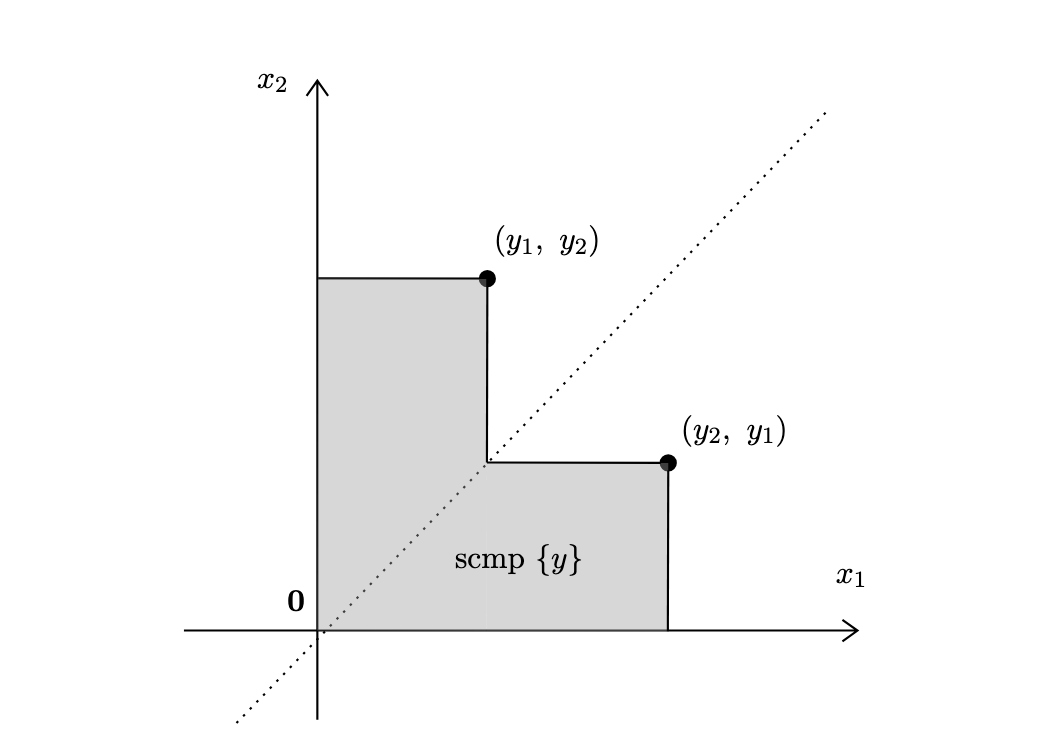}
    \caption{$\text{scmp} \{ y \}$ for some $y$ when $n= 2$}
    \label{fig_scomp}
\end{figure}

For a set $A \subset \mathbb{R}^n$, let $\text{cmp} \,  A$ denote the comprehensive hull of $A$. That is, $\text{cmp} \,A$ is the smallest comprehensive set including $A$, and can be written as 
\begin{equation*}
    \text{cmp} \,A = \{ x\in \mathbb{R}^n_{+} \mid \text{there exists $a\in A$ such that $a\geq x$} \}.
\end{equation*}
Similarly, for a set $A \subset \mathbb{R}^n$, the symmetric comprehensive hull of $A$ is denoted by $\text{scmp} \,A$, that is,
\begin{equation*}
    \text{scmp}\, A = \{ x\in \mathbb{R}^n_+ \mid \text{there exist $a\in A$ and  $\pi \in \Pi$ such that $a^\pi \geq x$} \}.
\end{equation*}
See Figure \ref{fig_scomp},  for an illustration of symmetric comprehensive hulls when $n = 2$. 
These two notations are used repeatedly throughout this paper.

%%%%%%%%%%%%%%%%%%%%%%%%%%%%%%%%%%%%%%%%%%%%%%%%%%%
\section{Rationalizability with the Normalization}
\label{sec_axioms}
%%%%%%%%%%%%%%%%%%%%%%%%%%%%%%%%%%%%%%%%%%%%%%%%%%%

Natural or reasonable properties for choice rules, referred to as \textit{axioms}, have been studied in the literature. 
This section introduces basic axioms and examines their implications. 
We begin with axioms of efficiency. 
The following two conditions have been widely studied. 

\vspace{3mm}
\begin{description}
    \item[\bf Strong Pareto.] For all $X\in \mathcal{P} $ and $x\in F(X)$, there is no $y \in X$ such that $y > x$.
    \item[\bf Weak Pareto.] For all $X\in \mathcal{P} $ and $x\in F(X)$, there is no $y \in X$ such that $y \gg x$. 
\end{description}
\vspace{3mm}
It is known that several well-known criteria (e.g., Rawlsian maximin choice rule) satisfy only 
 \textit{weak Pareto}. 
To accommodate these choice rules into our scope, we do not impose \textit{strong Pareto}. 
Also, note that solutions that satisfy only \textit{weak Pareto} sometimes discard all strictly efficient outcomes.  
However, it is difficult to justify ignoring all strictly Pareto efficient outcomes and choosing only weakly efficient ones. 
We then consider an intermediate condition: The following axiom requires that for all problems, all chosen outcomes should be weakly Pareto efficient and at least one of them should be strictly Pareto efficient.
\vspace{3mm}
\begin{description}
    \item[\bf Intermediate Pareto.] For all $X\in \mathcal{P} $ and $x\in F(X)$, there is no $y \in X$ such that $y \gg x$. Furthermore, for some $x' \in F(X)$, there is no $y' \in X$ such that $y' > x'$. 
\end{description}
\vspace{3mm}

The next axiom concerns the invariance property under individual-wise positive scalar multiplication.
This postulates that if agent $i$ makes an effort and as a result, achieves utility levels that are 
 $a_i$ times higher than in the original problem, then the social planner should respect his/her effort and give him/her $a_i$ times higher utility levels in the chosen outcomes. 
  In other words, it requires that the chosen outcomes be determined proportionally to the utility levels individuals can achieve.
\vspace{3mm}
\begin{description}
    \item[\bf Scale Invariance.] For all $X\in \mathcal{P} $ 
 and $a \in \mathbb{R}^n_{++}$, $F(a X) = a F(X)$.
\end{description}
\vspace{3mm}

We also impose the axiom of impartiality among individuals.

\vspace{3mm}
\begin{description}
    \item[\bf Anonymity.] For all symmetric problems $X\in \mathcal{P}$, if $x\in F(X)$, then $ x^\pi \in F(X)$ for all  $\pi \in \Pi$. 
\end{description}
\vspace{3mm}

We consider a weak axiom of choice invariance as well. 
When examining choice rules, axioms related to choice invariance have been studied (cf. \citealp{arrow1959rational,nash1950bargaining}). 
Roughly speaking, these axioms postulate that if a choice rule outcome in the original problem remains feasible in a smaller problem, then it should be chosen in the smaller problem. 
However, they have been criticized for requiring consistency even when the balance of bargaining power changes significantly.
To address this issue, our axiom requires contraction independence only when all individuals have equal abilities in both the original and the contracted problems. 
Let $\mathcal{P}^e$ be the set of problems $X$ where $b_i (X) = b_j (X)$ for all $i,j\in N$. 
The formal definition is as follows: 

\vspace{3mm}
\begin{description}
    \item[\bf Contraction Independence for Equal-Able Individuals (EAI).] For all $X, X'\in \mathcal{P}^e$ with $X'\subset X$, if $X' \cap F(X) \neq \emptyset$, then $F(X') = X' \cap F(X)$.\footnote{Compared with other weakenings of Nash's IIA, this axiom is weaker than the counterpart independence axiom of \citet{DUBRA2001131}, which defines choice rules as single-valued functions. A similar axiom was also considered in \citet{xu2006alternative}. } 
\end{description}
\vspace{3mm}

Axioms about continuity are also widely considered. The following requires that small changes in problems should not lead to large changes in chosen outcomes.
\vspace{3mm}
\begin{description} 
    \item[\bf Continuity.] For all $X\in \mathcal{P}$ and $x\in X$, if there exists $\{X^k \}_{k\in \mathbb{N}} \subset \mathcal{P}$ and $\{ x^k\}_{k\in \mathbb{N}}\subset \mathbb{R}^n$ such that (i)  $x^k \in F(X^k)$ for all $k\in \mathbb{N}$; (ii) $\{X^k \}_{k\in \mathbb{N}}$ converges to $X$ in the Hausdorff topology; (iii) $\{ x^k\}_{k\in \mathbb{N}}$ converges to $x$, then $x\in F(X)$.
\end{description}
\vspace{3mm}

By considering these basic axioms, we obtain necessary and sufficient conditions for choice rules to be ``rationalized" by some ability-normalized welfare function or ordering with several basic properties. 
Let $R$ be a weak order (i.e., a complete and transitive binary relation) over $\mathbb{R}^n_+$.\footnote{
We say that a binary relation $R$ over $\mathbb{R}^n_+$ is complete if for all $x, y\in \mathbb{R}^n_+$, $xR y$ or $yRx$; $R$ is transitive if for all $x, y, z\in \mathbb{R}^n_+$, $xR y$ and $yRz$ imply $xRz$. 
} The symmetric part and asymmetric part are denoted by $I$ and $P$, respectively. 
A choice rule $F$ is said to be \textbf{\textit{rationalized by a normalized welfare ordering}} $R$ if  for all $X\in \mathcal{P}$, 
\begin{equation*}
    F (X) = \qty{ x\in X ~ \bigg| ~ \text{$\not\exists y\in X$ s.t. $\qty( {y_1\over b_1(X)}, \cdots, {y_n\over b_n(X)} )  P \qty( {x_1\over b_1(X)}, \cdots, {x_n\over b_n(X)} ) $} }. 
\end{equation*}
We say that  an ordering $R$ is \textit{weakly monotone} if $x P y$ for all $x, y \in \mathbb{R}^n_+$ with $x \gg y$, and $x R y$ for all $x, y \in \mathbb{R}^n_+$ with $x \geq y$.
An ordering $R$ is \textit{symmetric} if $x I x^\pi$ for all $x\in \mathbb{R}^n_+$ and all  $\pi \in \Pi$.
An ordering $R$ is \textit{homogeneous} if $xRy \iff \alpha x R \alpha y$ for all $x, y \in \mathbb{R}^n_+$ and all $\alpha \in \mathbb{R}_{++}$.

By imposing the above axioms except for \textit{continuity}, we can characterize the choice rules that are rationalized by some normalized welfare ordering with the properties introduced above. 

\vspace{3mm}
\begin{theorem}
\label{thm_rationalordering}
     A choice rule $F$ satisfies \textit{intermediate Pareto}, \textit{scale invariance}, \textit{anonymity}, and \textit{contraction independence for EAI} if and only if there exists a weakly monotone, symmetric and homogeneous weak order $R$ over $\mathbb{R}^n_+$ such that $F$ is  is rationalized by the normalized welfare ordering $R$.  
\end{theorem}
\vspace{3mm}

By imposing \textit{continuity} additionally, a choice rule can be represented by some continuous function. 
A choice rule $F$ is said to be  \textbf{\textit{rationalized by a normalized welfare function}}  $W:\mathbb{R}^n_{+} \rightarrow \mathbb{R}$ if for all $X\in \mathcal{P}$, 
\begin{equation*}
    F (X) = \argmax_{x\in X} ~  W \qty( {x_1 \over b_1 (X) }, {x_2 \over b_2 (X) }, \cdots, {x_n \over b_n (X) } ).
\end{equation*}
Let $\mathcal{D}\in \{ \mathbb{R}^n_+,  \mathbb{R}^n \} $. We say that a function $W:\mathcal{D}\rightarrow \mathbb{R}$ is \textit{weakly monotone} if $W(x) > W(y)$ for all $x, y \in\mathcal{D}$ with $x \gg y$.
A function $W:\mathcal{D}\rightarrow \mathbb{R}$ is \textit{symmetric} if $W(x) = W(x^\pi)$ for all $x\in\mathcal{D}$ and all  $\pi \in \Pi$.
A function $W:\mathcal{D}\rightarrow \mathbb{R}$ is \textit{homogeneous} if $W(\alpha  x) = \alpha W(x)$ for all $x\in\mathcal{D}$ and all $\alpha \in \mathbb{R}_{++}$.
Then our second characterization result is as follows: 

\vspace{3mm}
\begin{theorem}
\label{thm_rational}
    A choice rule $F$ satisfies \textit{intermediate Pareto}, \textit{scale invariance}, \textit{anonymity}, \textit{contraction independence for EAI}, and \textit{continuity} 
    if and only if there exists a weakly monotone, symmetric, homogeneous and continuous function  $W:\mathbb{R}^n_{+} \rightarrow \mathbb{R}$ such that $F$ is  is rationalized by the normalized welfare function $W$. 
\end{theorem}
\vspace{4mm}

These results provide axiomatic foundations for the idea that choice rules should be "rationalized" by some welfare criteria with  normalized by individuals’ abilities.
Although necessary and sufficient conditions for the rationalizability of choice rules have been studied in the literature, they have examined only welfare relations or functions that can be defined independently of what vectors are feasible (e.g., \citealp{bossert1994rational,denicolo2000nash,ok1999revealed,peters1991independence,sanchez2000rationality,xu2013rationality}). 
Since the set of feasible utility vectors reflects the individuals' productivity, it is natural to evaluate each outcome depending on the shape of the problems. 
Theorem \ref{thm_rationalordering} and \ref{thm_rational} characterize the choice rules that are rationalized by orderings or functions with a particular form of normalization.

These results are useful when considering choice rules that choose outcomes proportional to the levels of utility individuals can achieve. Indeed, Theorem \ref{thm_rational} plays an important role in the proofs of results in Section \ref{sec_specialcases} as well: They can be derived simply by combining this general characterization with existing results in social choice theory. 

Also, it should be noted that a companion paper (\citealp{nakamura2025weakindependenceirrelevantalternatives}) provides an alternative characterization of the solutions considered in Theorem \ref{thm_rational}. 
This paper shows that these solutions can be represented using the normalized Nash products even though they do not necessarily satisfy Nash's IIA. Since the original Nash solution can be rewritten as assigning the maximizer of the normalized Nash products to each problem, this result can be considered as claiming that evaluations based on products of utility levels do not solely stem from Nash's IIA.

%%%%%%%%%%%%%%%%%%%%%%%%%%%%%%%%%%%%%%%%%%%%%%%%%%%
\section{The Relative Fair Choice Rules}
\label{sec_modifiedKS}
%%%%%%%%%%%%%%%%%%%%%%%%%%%%%%%%%%%%%%%%%%%%%%%%%%%

By considering two additional axioms, we characterize a new class of choice rules that can be viewed as  combinations of three key principles: distribution according to individuals' abilities, utilitarianism, and  egalitarianism.

One is an axiom of invariance with respect to the addition of a common utility. 
If all individuals exogenously gain the same amount of utility, then it is natural to require that the chosen utility vector changes by the same amount.
The following requires this invariance property only when all individuals have equal abilities. 
\vspace{3mm}
\begin{description}
    \item[\bf Equal Addition Independence for Equal-Able Individuals (EAI).] For all $X \in \mathcal{P}^e$, $x \in X$ and $\alpha \in \mathbb{R}_{++}$,  
    $x\in F(X)$ if and only if $x + \alpha \mathbf{1} \in F( \text{cmp} (X + \alpha \mathbf{1} ))$. 
\end{description}
\vspace{3mm}
Compared with other invariance properties with respect to shifts of problems (e.g., \citealp{blackorby1994generalized,hinojosa2008inequality,ok1999revealed,ok2000choquet,thomson1981nash}), this axiom is quite mild.  

The other axiom is about compromisability. It requires that if  individuals have equal abilities, then compromises of chosen outcomes should also be chosen:
More precisely, for any pair of chosen outcomes $x$ and $y$ and $\alpha\in[0,1]$,  either the convex combination $\alpha x + (1 - \alpha) y$ itself or outcomes that Pareto-dominates $\alpha x + (1 - \alpha) y$ should be chosen.
\vspace{3mm}
\begin{description}
    \item[\bf Compromisability for Equal-Able Individuals (EAI).] For all $X\in \mathcal{P}^e$, $x, y \in F(X)$ and $\alpha \in [0,1]$, if $\alpha x + (1 - \alpha) y \in X$, then there exists $z\in F(X)$ such that $z \geq \alpha x + (1 - \alpha) y $. 
\end{description}
\vspace{3mm}
\citet{ok1999revealed,ok2000choquet} introduced an axiom with a similar motivation.

By using these two axioms in addition to the axioms in Theorem \ref{thm_rational}, we characterize a new class of choice rules. Let $\Delta$ be the  set of weights over the individuals, i.e., $\Delta = \{ w\in \mathbb{R}_+^n \mid\sum_{i\in N} w_i = 1\} $. 

\vspace{3mm}
\begin{theorem}
\label{thm_main}
    A choice rule $F$ satisfies \textit{intermediate Pareto}, \textit{scale invariance}, \textit{anonymity}, \textit{contraction independence for EAI}, \textit{continuity},  \textit{equal addition independence for EAI}, and \textit{compromisability for EAI}
    if and only if
    there exists a nonempty symmetric closed convex set $\mathcal{W}\subset \Delta$ such that for all $X\in \mathcal{P}$, 
\begin{equation}
\label{eq_modiKS}
    F (X) = \argmax_{x\in X} \min_{w\in \mathcal{W} } \sum_{i\in N} w_i {x_i \over b_i (X)}. 
\end{equation}
\end{theorem}
\vspace{4mm}
\noindent
We provide a proof of Theorem \ref{thm_main} and verify the independence of the axioms in Appendix.

We refer to the choice rule represented by \eqref{eq_modiKS} as the \textbf{\textit{relative fair choice rule}} \textit{associated with the weight set} $\mathcal{W}$.
These choice rules can be interpreted as decision-making rules of an inequality-averse social planner who has in mind a symmetric set of plausible weights over the individuals. 
She evaluates each utility vector as follows:  After normalizing the individuals' utilities by their abilities, she computes the weighted sum of the normalized utility levels with respect to each plausible weight  (in a \textbf{utilitarian} way).
Then, to give higher weights to disadvantaged individuals, she chooses the smallest value as the utility vector evaluation (in an \textbf{egalitarian} way).
The weight set reflects the degree of inequality aversion of the social planner: The larger the weight set is, the more inequality-averse she is.  
In the next two sections, we will further examine this class of choice rules.

%%%%%%%%%%%%%%%%%%%%%%%%%%%%%%%%%%%%%%%%
\section{Special Cases}
\label{sec_specialcases}
%%%%%%%%%%%%%%%%%%%%%%%%%%%%%%%%%%%%%%%%

The class of relative fair choice rules encompasses many interesting subclasses of choice rules as special cases. 
This section examines several examples and provides axiomatic foundations for the extreme cases. 

%%%%%%%%%%%%%%%%%%%%%%%%%%%%%%%%%%%%%%%%
\subsection{The Relative Maximin Choice Rule}
\label{subsec_Rmaximin}
%%%%%%%%%%%%%%%%%%%%%%%%%%%%%%%%%%%%%%%%

When $\mathcal{W}$ is the entire set, i.e., $\mathcal{W} =\Delta$, then the relative fair choice rule associated with the weight set $\mathcal{W}$ can be written as for all $X\in \mathcal{P}$, 
\begin{equation}
\label{eq_modiKS_min}
    F (X) = \argmax_{x\in X} \min_{i\in N} {x_i \over b_i (X)}. 
\end{equation}
This can be interpreted as the choice rule of an inequality-averse social planner who evaluates each utility vector using Rawls' (\citeyear{Rawls1971-JOHATO-12}) maximin principle after normalizing the individuals' utilities by their abilities. 
We refer to this choice rule as the \textit{relative maximin choice rule}. 

This choice rule is very similar to the choice rule introduced \citet{kalai1975other} in axiomatic bargaining theory. 
A choice rule $F$ is called the \textit{KS choice rule}  if it  assigns to every problem $X\in \mathcal{P}$ the point $x \in X$ such that $x_i/b_i(X) = x_j / b_j (X)$ for all $i,j\in N$ and there is no $y\in X$ with $y\gg x$.
That is, the KS choice rule selects the weakly Pareto efficient outcome proportional to individuals' abilities. 
For each $X$, the KS choice rule can be interpreted as maximizing the welfare level 
\begin{equation*}
     \min_{i\in N} {x_i \over b_i (X)}
\end{equation*}
over $x\in X$ and breaking ties by choosing the outcome $x$ such that $x_i/b_i(X) = x_j / b_j (X)$ for all $i,j\in N$. 

The only difference between them is whether ties are broken after maximizing the objective function in \eqref{eq_modiKS_min}. It is worth noting that these two choice rules are equivalent if the problems are assumed to be convex. 
For a more detail discussion about the logical relationship between them, see Section \ref{sec_comparison}. 

The relative maximin choice rule can be characterized using a variant of the equity principle formalized by \citet{hammond1976equity} in the context of social welfare orderings. 
This equity axiom requires that in any two-individual conflict,  the least advantaged one should always be prioritized. 
The following axiom postulates this egalitarian principle only when the individuals' abilities are equal. 
\vspace{3mm}
\begin{description}
    \item[Hammond Equity for Equal-Able Individuals (EAI).] Let $x, y\in \mathbb{R}^n_+$ be such that  \begin{align*}
        \begin{array}{cc}
    x_i < y_i < y_j < x_j, & \text{for some }i, j \in N, \\
   x_k = y_k, & \text{for all }k\in N \backslash\{ i, j \}. 
    \end{array}
    \end{align*}
    Then for all $X\in \mathcal{P}^e$, $x\in F(X)$ and $y\in X$ imply $y\in F(X)$. 
\end{description}
\vspace{3mm}
The next result shows that the relative maximin choice rule can be obtained from the axioms in Theorem \ref{thm_rational} and \textit{Hammond equity for EAI}. 

\vspace{3mm}
\begin{proposition}
\label{prop_relmaximin}
    A choice rule $F$ satisfies \textit{intermediate Pareto}, \textit{scale invariance}, \textit{anonymity}, \textit{contraction independence for EAI}, \textit{continuity}, and \textit{Hammond equity for EAI}
    if and only if it is the relative maximin choice rule. 
\end{proposition}

%%%%%%%%%%%%%%%%%%%%%%%%%%%%%%%%%%%%%%%%
\subsection{The Relative Utilitarian Choice Rule}
\label{subsec_Rutil}
%%%%%%%%%%%%%%%%%%%%%%%%%%%%%%%%%%%%%%%%

When $\mathcal{W}$ is a singleton, i.e., $\mathcal{W} = \{ (1/n, 1/n, \cdots, 1/n ) \}$, then the relative fair choice rule associated with the weight set $\mathcal{W}$ can be written as for all $X\in \mathcal{P}$, 
\begin{equation}
    F (X) = \argmax_{x\in X} \sum_{i\in N}{x_i \over b_i (X)}. 
\end{equation}
This choice rule chooses the maximizers of the sum of utility levels normalized by each individual's abilities.  
We refer to it as the \textit{relative utilitarian choice rule}. 
Similar criteria have been studied in various contexts (e.g., \citealp{dhillon1999relative,karni1998impartiality,segal2000let,sprumont2019relative}). 
In axiomatic bargaining theory, the relative utilitarian choice rule has been axiomatized by \citet{baris2018timing} and \citet{peitler2023rational},  and examined by \citet{cao1982preference}, \citet{pivato2009twofold}, and \citet{rachmilevitch2015nash,rachmilevitch2016egalitarian}.

By using Theorem \ref{thm_rational}, we provide a new axiomatic foundation for this rule in the cases where $n\geq 3$. 
Consider a new axiom that postulates the independence of unconcerned individuals when all individuals have equal abilities.
For all nonempty set $M\subset N$ and $x, y\in \mathbb{R}^n$, let $(x_{M}, y_{M\backslash N})$ be the utility vector $z$ such that $z_i = x_i$ for all $i\in M$ and  $z_i = y_i$ for all $i\in N \backslash M$.

Suppose that  $(x_M , x_{N\backslash M}) \in F(X)$ and $(y_M, x_{N\backslash M}) \notin F(X)$ for some $X\in \mathcal{P}^e$. 
This can be interpreted as indicating that the chosen outcome $(x_M , x_{N\backslash M})$ is more desirable than the not-chosen one $(y_M, x_{N\backslash M})$. 
If we admit the principle of independence of unconcerned individuals---which requires that, when we compare two utility vectors, individuals achieving the same utility levels in both vectors do not influence which vector is preferred---for individuals with equal abilities, then $(x_M , y_{N\backslash M})$ should be more desirable than  $(y_M, y_{N\backslash M})$.  
That is, in any problem where all individuals have equal abilities, if both $(x_M , y_{N\backslash M})$ and $(y_M, y_{N\backslash M})$ are feasible, then $(y_M, y_{N\backslash M})$ should not be chosen. 

\vspace{3mm}
\begin{description}
    \item[\bf Separability for Equal-Able Individuals (EAI).]
    Let $M$ be a nonempty subset of $N$. 
    For all $x, y\in \mathbb{R}^n_+ \backslash \{ \mathbf{0} \}$ and $X, X' \in \mathcal{P}^e$, if $(x_M , x_{N\backslash M}) \in F(X)$, $(y_M, x_{N\backslash M}) \notin F(X)$ and $(x_M, y_{N\backslash M}) \in X'$, then $(y_M, y_{N\backslash M}) \notin F(X')$. 
\end{description}
\vspace{3mm}
%===== explain

\begin{proposition}
\label{prop_relutil}
    Suppose that $n\geq 3$. A choice rule $F$ satisfies \textit{intermediate Pareto}, \textit{scale invariance}, \textit{anonymity}, \textit{contraction independence for EAI}, \textit{continuity}, \textit{equal addition independence for EAI}, and \textit{separability for EAI}
    if and only if it is the relative utilitarian choice rule. 
\end{proposition}

This result cannot be extended to the case where $n = 2$. To see this, consider a choice rule $F$ defined as for all $X\in \mathcal{P}$,
\begin{equation*}
    F(X) = \max_{x\in X} ~ \qty( \alpha_1 \min_{i\in \{1,2 \}} {x_i \over b_i(X)} + \alpha_2  \max_{i\in \{1,2 \}} {x_i \over b_i(X)} ), 
\end{equation*}
where $\alpha_1, \alpha_2 > 0$ and $\alpha_1 \neq \alpha_2$. This choice rule satisfies all of the axioms in Proposition \ref{prop_relutil}. 
For characterizations of the relative utilitarian choice rule including the case where $n=2$, see \citet{baris2018timing} and \citet{peitler2023rational}.

%%%%%%%%%%%%%%%%%%%%%%%%%%%%%%%%%%%%%%%%
\subsection{Other Special Cases}
\label{subsec_otherspec}
%%%%%%%%%%%%%%%%%%%%%%%%%%%%%%%%%%%%%%%%

The class of relative fair choice rules includes other interesting choice rules as special cases. 
This section discusses some of them without providing axiomatic foundations for them. 

When $\mathcal{W} = \alpha \{ 1/n, \cdots, 1/n  \} + (1- \alpha) \Delta$ for some $\alpha \in [0,1]$, the relative fair choice rule associated with the weight set $\mathcal{W}$ maximizes a convex combinations of the objective functions of the relative maximin choice rule and the relative utilitarian choice rule, that is, for all $X\in \mathcal{P}$, 
\begin{equation*}
    F(X) = \argmax_{x\in X} ~ \qty( \alpha \sum_{i\in N} {x_i \over b_i(X)} + (1-\alpha)\min_{i\in N} {x_i \over b_i (X)} ). 
\end{equation*}
This class of choice rules can be viewed as the simplest compromise between the two extreme cases.
\citet{rachmilevitch2015nash,rachmilevitch2016egalitarian} considered a choice rule within this class to examine the properties of the Nash choice rule. 
In the literature on social choice theory, \citet{roberts1980interpersonal} introduced convex combinations of the utilitarian social welfare function and the maximin social welfare function (without any normalization)  and \citet{BK2020ET} first characterized them. 

We consider a more generalized class of choice rules. 
 When $\mathcal{W}$ is  the convex hull of the set $\{ a\in \Delta \mid a = w^\pi ~ \text{for some  $\pi \in \Pi$}\}$ for some $w\in \Delta$, the relative fair choice rule associated with the weight set $\mathcal{W}$ can be written as follows: For all $X\in \mathcal{P}$, 
\begin{equation*}
    F(X) = \argmax_{x\in X} \sum_{i \in N} w_{(i)} \widetilde{x}_{(i)}, 
\end{equation*}
where $\widetilde{x}$ is defined as $\widetilde{x}_i = x_i / b_i(X)$ for all $x\in X$ and for all $i\in N$. 
It maximizes the generalized Gini welfare function (\citealp{weymark1981generalized}) associated with $w$ after normalizing the individuals' utility levels by their  abilities. 
\citet{blackorby1994generalized} and \citet{ok1999revealed,ok2000choquet} examined the generalized Gini choice rules, although they did not consider any normalization. 

Furthermore, the relative fair choice rules include 
choice rules $F$ written as for some $\theta\in[0,1]$,  
\begin{equation}
\label{eq_meanvar}
    F(X) = \argmax_{x\in X} ~ \qty{ {1 \over n}\sum_{i\in N} {x_i \over b_i(X)} -  \theta~  \text{SD} \qty( {x_1 \over b_1 (X)} , \cdots, {x_n \over b_n (X)} ) },
\end{equation}
where $\text{SD}:\mathbb{R}^n\rightarrow \mathbb{R}$ is a function defined as for all $y \in \mathbb{R}^n$, 
\begin{equation*}
    \text{SD} (y) =  \qty{ {1 \over n}\sum_{i\in N} \qty(y_i - {1 \over n}\sum_{j \in N} y_i )^2 }^{1 \over 2}. 
\end{equation*}
The choice rule in \eqref{eq_meanvar} evaluates each $x \in X$ based on the mean and variance of the normalized vector.
The variance term acts as a penalty for selecting unequal outcomes. 
% \footnote{
% Note that \citet{hinojosa2008inequality} considers solutions different form the relative fair solutions in two points : (i) Their solution does not normalize the individuals' utility level and (ii) the set of plausible weight $\mathcal{W}$ is a finite set. 
% Of course, axioms supporting each solution concepts are quite different. 
% Furthermore, the classes of solution they include is also different other than whether normalization is done or not. 
% For example, solutions represented as \eqref{eq_meanvar} are  are not included if the weight set is finite.}
\citet{roberts1980interpersonal} proposed the corresponding social welfare orderings without normalization.

\begin{remark}
    If $\theta > 1$ in \eqref{eq_meanvar}, then it violates \textit{intermediate Pareto}, that is, it is not a relative fair choice rule. 
    To see this, consider the two-person case and let $W:\mathbb{R}^2_+ \rightarrow \mathbb{R}$ be a function such that for all $x\in \mathbb{R}^n$, 
    \begin{equation*}
        W(x)
        = 
        {1 \over 2} (x_1 + x_2)-  \theta~  \qty{ {1 \over 2} \qty( x_1 - {1 \over 2} (x_1 + x_2) )^2 +  {1 \over 2} \qty( x_2 - {1 \over 2} (x_1 + x_2) )^2 }^{1\over 2}. 
    \end{equation*}
    It is sufficient to show that this function is not weakly monotone if $\theta > 1$. Let $\delta \in (1, \theta)$. 
    The function $W$ can be rewritten as
    \begin{equation*}
        W(x)
        = 
        {1 \over 2} (x_1 + x_2)-  
        \theta~   {1 \over 2} | x_2 - x_1 |. 
    \end{equation*}
    Since $ W(3 - \delta , 3- \delta)  = 3-\delta >  3 - \theta = W(2,4)$, $W$ is not weakly monotone. 
\end{remark}

In choice rules defined as \eqref{eq_meanvar},the penalty term is measured using the distance induced by the Euclidean norm (i.e., the simple Euclidean distance). Similarly, other symmetric norms can also be considered.\footnote{
Formally, we say that a function $\| \cdot \| : \mathbb{R}^n \rightarrow \mathbb{R}$ is a norm if it satisfies the following conditions: 
\begin{itemize}
    \item $\| x \| \geq 0$ for all $x\in \mathbb{R}^n$,  and $\| x \| = 0$ if and only if $x = \mathbf{0}$.  
    \item $\| \alpha x \| = |\alpha| \| x \|$ for all $\alpha \in \mathbb{R}$ and $x\in \mathbb{R}^n$. 
    \item $\| x + y \| \leq \| x \| + \| y \|$ for all $x, y\in \mathbb{R}^n$. 
\end{itemize}
} 
This allows for various ways of representing penalties for inequalities.
Given a symmetric norm $\| \cdot \| : \mathbb{R}^n \rightarrow \mathbb{R}$, consider a choice rule defined as for all $X\in \mathcal{P}$, 
\begin{equation}
\label{eq_meannorm}
    F(X) = \argmax_{x\in X} ~ \qty{ {1 \over n}\sum_{i\in N} \widetilde{x}_i -  \Bigg\| \widetilde{x} - \qty( {1 \over n}\sum_{i\in N} \widetilde{x}_i )  \mathbf{1} \Bigg\| }, 
\end{equation}
where $\widetilde{x} \in \mathbb{R}^n$  such that $\widetilde{x}_i = x_i / b_i(X)$ for all $x\in X$ and for all $i\in N$. 
If 
\begin{equation}
\label{eq_meannorm_cond}
     {1 \over n}\sum_{i\in N} x_i -  \Bigg\| x - \qty( {1 \over n}\sum_{i\in N} x_i )  \mathbf{1} \Bigg\|  > 0 
\end{equation}
for all $x\in \mathbb{R}^n_{++}$, then the objective function in \eqref{eq_meannorm} is  symmetric, homogeneous, quasiconcave, weakly monotone, and continuous.\footnote{
To see the quasiconcavity, let $W: \mathbb{R}^n_+\rightarrow \mathbb{R}$ be the function defined as for all $x\in \mathbb{R}^n$, 
\begin{equation*}
    W(x) =  {1 \over n}\sum_{i\in N} x_i -  \Bigg\| x - \qty( {1 \over n}\sum_{i\in N} x_i )  \mathbf{1} \Bigg\| . 
\end{equation*}
Let $x, y\in \mathbb{R}^n$ and $\alpha \in [0,1]$. Without loss of generality, we assume $W(x) \geq W(y)$. Then, we have 
\begin{align*}
    &W(\alpha x + (1 - \alpha ) y) \\
    &= {1 \over n}\sum_{i\in N} (\alpha x_i + (1 - \alpha ) y_i ) -  \Bigg\| \alpha x + (1 - \alpha ) y - \qty( {1 \over n}\sum_{i\in N} (\alpha x_i + (1 - \alpha ) y_i )  )  \mathbf{1} \Bigg\| \\
    &\geq {1 \over n}\sum_{i\in N} (\alpha x_i + (1 - \alpha ) y_i ) -  \Bigg\| \alpha x -  \qty( {1 \over n}\sum_{i\in N}\alpha x_i   )  \mathbf{1}  \Bigg\|
    - \Bigg\| (1 - \alpha ) y - \qty( {1 \over n}\sum_{i\in N}  (1 - \alpha ) y_i  )  \mathbf{1} \Bigg\| \\
    &= \alpha \qty( \sum_{i\in N} x_i  - \Bigg\| x -  \qty( {1 \over n}\sum_{i\in N}x_i )  \mathbf{1}  \Bigg\| ) +  (1 - \alpha) \qty( \sum_{i\in N} y_i  - \Bigg\| y -  \qty( {1 \over n}\sum_{i\in N}y_i )  \mathbf{1}  \Bigg\| ) \\
    &= \alpha W(x) + (1 -\alpha) W(y)\\
    &\geq W(y). 
\end{align*}
Therefore, the function $W$ is quasiconcave. 
} 
Therefore, when \eqref{eq_meannorm_cond} holds for all $x\in \mathbb{R}^n_{++}$, choice rule \eqref{eq_meannorm} is a relative fair choice rule.\footnote{
If the set of plausible weight $\mathcal{W}$ is a finite set (e.g., \citealp{hinojosa2008inequality}), choice rules in \eqref{eq_meannorm} (including \eqref{eq_meanvar}) cannot be represented as the relative fair choice rules in general. 
}
(For details, see the proof of Theorem \ref{thm_main}, where we construct the weight set from a given symmetric, homogeneous, quasiconcave, weakly monotone, and continuous function.)

\begin{remark}
    Note that \eqref{eq_meannorm_cond} is a necessary and sufficient condition for the objective function in \eqref{eq_meannorm} to be weakly monotone. 
    This condition is always satisfied, for example, when $\| \mathbf{1} \| \leq 1$. 
\end{remark}

%%%%%%%%%%%%%%%%%%%%%%%%%%%%%%%%%%%%%%%%%%%%%%%%%%%
\section{Comparison with the Kalai-Smorodinsky Choice Rule}
\label{sec_comparison}
%%%%%%%%%%%%%%%%%%%%%%%%%%%%%%%%%%%%%%%%%%%%%%%%%%%

As discussed in Section \ref{subsec_Rmaximin}, a special case of the relative fair choice rules is closely related to the KS choice rule,  one of the dominant solution concepts in axiomatic bargaining theory.
This section examines the relationship between the KS choice rule and the relative fair choice rules at the axiomatic level.

There are two differences between these choice rule concepts.
The first one is about efficiency: The KS choice rule satisfies \textit{weak Pareto} but not \textit{intermediate Pareto}. 
This means that it sometimes chooses only outcomes that are inefficient in terms of \textit{strong Pareto}; in other words, the KS choice rule sometimes gives up Pareto improvements. 
In contrast, the relative fair choice rule satisfies \textit{intermediate Pareto} and furthermore, if $\mathcal{W}$ is in the interior of $\Delta$, it satisfies \textit{strong Pareto}.

Another difference between them is the axioms of impartiality. 
Although both of them satisfy \textit{anonymity},
only the KS choice rule satisfies the following stronger condition related to impartiality due to \citet{xu2006alternative}: 
\vspace{3mm}
\begin{description}
    \item[\bf Strong Symmetry.] For all symmetric problems $X\in \mathcal{P}$, if $x\in F(X)$, then $x_1 = x_2 = \cdots = x_n$.  
\end{description}
\vspace{3mm}
This requires that for all symmetric problems, choice rules should choose outcomes where all people achieve the same utility level. It postulates not only impartiality but also distributive equity. 

Indeed, the KS choice rule can be characterized using axioms including \textit{weak Pareto} and \textit{strong symmetry}, and it satisfies all the axioms in Theorem 1 except for \textit{intermediate Pareto}. 
The first part can be shown by just a minor modification of  Xu and Yoshihara's (\citeyear{xu2006alternative}) proof of Theorem 3.\footnote{Note that \citet{xu2006alternative} used a different version of weak contraction independence.} For details, see Appendix. 

\vspace{3mm}
\begin{proposition}
\label{prop_originalKS}
    A choice rule satisfies \textit{weak Pareto}, \textit{scale invariance}, \textit{strong symmetry}, and \textit{contraction independence for EAI}
    if and only if it is the KS choice rule. 
    
    Furthermore, the KS choice rule satisfies \textit{anonymity}, \textit{continuity}, \textit{equal addition independence for EAI}, and \textit{compromisability for EAI}, that is, all the axioms in Theorem 1 except for \textit{intermediate Pareto}. 
\end{proposition}
\vspace{3mm}

Moreover, in the two-person case, by weakening \textit{intermediate Pareto} to \textit{weak Pareto} in Theorem 1, the KS choice rule and the relative fair choice rules can be jointly characterized. 

\vspace{3mm}
\begin{theorem}
\label{thm_joint}
    Suppose that $n = 2$. A choice rule $F$ satisfies \textit{weak Pareto}, \textit{scale invariance}, \textit{anonymity}, \textit{contraction independence for EAI}, \textit{continuity},  \textit{equal addition independence for EAI}, and \textit{compromisability for EAI} if and only if it is either the KS choice rule or a relative fair choice rule. 
\end{theorem}
\vspace{3mm}

These results suggest that these two choice rule concepts share many desirable properties and that the relative fair choice rules are natural modifications of the KS choice rule at the axiomatic level as well. 

Our proof in Theorem \ref{thm_joint} cannot be applied to the cases where $n\geq 3$. 
In Appendix, after providing a proof, we briefly explain why that proof is not valid when $n\geq 3$. 
 Whether this theorem can be generalized remains an open question for future work. 

%%%%%%%%%%%%%%%%%%%%%%%%%%%%%%%%%%%%%%%%%%%%%%%%%%%
\section{Related Literature}
\label{sec_literature}
%%%%%%%%%%%%%%%%%%%%%%%%%%%%%%%%%%%%%%%%%%%%%%%%%%%

This section briefly discusses the related literature. 

We have considered choice rules reflecting  differences in individuals' abilities. 
In the literature, fair allocation rules under unequal skills have been widely studied (e.g., \citealp{Fleurbaey2012Book}). 
Most of these studies focused on how to compensate for individuals' unequal skills, that is, examined the situations where individuals are not responsible for their skills. 
However, when equality of opportunity is achieved, it may be undesirable to fully compensate for their disadvantages. 
Instead, one could argue that these differences should be reflected in the chosen distribution, as they arise from the effort individuals have invested.
We have provided guidance for evaluating social outcomes in these situations, together with theoretical foundations underlying them. 

In the bargaining problem, \textit{scale invariance} has been interpreted as an axiom postulating the independence of utility representations (e.g., \citealp{nash1950bargaining}). 
Under this interpretation, interpersonal comparisons of utility level are entirely prohibited. On the other hand, we assume that utility levels and gains are comparable between individuals. 
In this setup, \textit{scale invariance} becomes an axiom  that determines how responsible each individual is for their attainable utility levels.
\citet{yoshihara2003characterizations} considered the bargaining problem in the market economy. In that paper, \textit{scale invariance} was used in the same way.

% As our first result, several papers have identified a set of axioms, including variants of Nash's IIA, for a necessary and sufficient condition for choice rules to be written as the maximal points of welfare functions or orderings without rescaling (\citealp{bossert1994rational,denicolo2000nash,ok1999revealed,peters1991independence,sanchez2000rationality,xu2013rationality}).
% However, since the  shape of the feasible sets reflects the productivity of the agents, it is natural to think that the shape of the problems should affect the chosen outcomes.
% Our first result provides an axiomatic foundation for this idea. 

Several papers suggested the Nash choice rule as a compromise between the relative maximin choice rule and the relative utilitarian choice rule by examining the geometric properties of chosen outcomes (e.g., \citealp{cao1982preference,rachmilevitch2015nash,rachmilevitch2016egalitarian}). 
However, the product of the individuals' utility levels, which the Nash choice rule maximizes, is difficult to interpret from a normative perspective. 
Compared to the Nash choice rule, the class of choice rules studied in this paper can be clearly understood  as a compromise of these two extreme choice rule concepts. 

Although not normalized, compromises between the egalitarian choice rule (\citealp{Kalai1977Econometrica}) and the utilitarian choice rule (\citealp{myerson1977two}) have been considered.
\citet{blackorby1994generalized} and \citet{ok1999revealed} axiomatized a class of choice rules corresponding to the generalized Gini social welfare functions (\citealp{weymark1981generalized}). 
\citet{ok1999revealed,ok2000choquet} proposed and characterized a more general class of choice rules related to the Choquet integral with monotonic capacity (\citealp{schmeidler1986integral,schmeidler1989subjective}).   
\citet{hinojosa2008inequality} axiomatized a new class of choice rules with finitely many plausible weights. Their choice rule evaluates each utility vector by computing the weighted sum of utility levels for each weight and picking the smallest one. 
In comparison, the relative fair choice rules normalize the individual's utility before computing the weighted sum and allow infinitely many plausible  weights in mind. 
Due to this generality, the class of relative fair choice rules includes various intuitive criteria that cannot be represented when the weight set is finite. (For example, see Section \ref{subsec_otherspec}.)
Moreover,  the relative fair choice rules are supported by a set of axioms that are easier to accept. 
Especially, they do not rely on controversial Nash's IIA, and our new axiom of invariance with respect to shifts is much milder than the counterpart in \citet{hinojosa2008inequality}.

See also a series of studies by \citet{rachmilevitch2015nash,rachmilevitch2016egalitarian,rachmilevitch2019egalitarianism,rachmilevitch2023nash,rachmilevitch2024nash}, which examine  the Nash choice rule as a compromise between the egalitarian choice rule and the utilitarian choice rule, without the normalization.

%%%%%%%%%%%%%%%%%%%%%%%%%%%%%%%%%%%%%%%%%%%%%%%%%%%
\section{Discussion}
\label{sec_discussion}
%%%%%%%%%%%%%%%%%%%%%%%%%%%%%%%%%%%%%%%%%%%%%%%%%%%

In this paper, we have proposed choice rules with appealing properties as social planners' choice rules when individuals are responsible for their abilities.  
First, we have identified the necessary and sufficient conditions for choice rules to be represented as the maximizers of an ability-normalized welfare function or ordering over individuals' utility levels (Theorem \ref{thm_rationalordering} and \ref{thm_rational}). 
Our general results have the potential to expand the scope of axiomatic bargaining theory by connecting it with social choice theory. 
Indeed, using these general results in conjunction with those from social choice theory, we provide new characterizations of the relative maximin and utilitarian choice rules (Proposition \ref{prop_relmaximin} and \ref{prop_relutil}).  

Next, we have proposed and characterized a new class of choice rules, which we call the relative fair choice rules, using the additional two axioms (Theorem \ref{thm_main}). 
These choice rules concern distribution according to abilities, efficiency, and equity. 
Furthermore, we have shown that this new class of choice rules is a natural modification of the  KS choice rule at the axiomatic level as well (Proposition \ref{prop_originalKS} and Theorem \ref{thm_joint}). 

To conclude this paper, we offer several remarks on future work that have not yet been discussed.

\begin{itemize}
    \item The relative fair choice rules evaluate utility vectors in a manner similar to the Maxmin Expected Utility models (\citealp{gilboa1989maxmin}), a class of decision rules under uncertainty. 
    Using the techniques developed in this paper, it would be possible to consider a menu-dependent version of Maxmin Expected Utility preferences. 
    In particular, Theorem \ref{thm_rational} would be useful for examining menu-dependent criteria under uncertainty.  

    \item The class of choice rules characterized in Theorem \ref{thm_rational} includes the Nash choice rule (\citealp{nash1950bargaining}). 
    It would be interesting to provide a new characterization of the Nash choice rule using this result without relying on the original IIA. 
    Since this is beyond the scope of this paper, we leave it to future work.

    \item We have characterized choice rules where individuals are fully responsible for their abilities. 
    In the literature, non-responsible or fully responsible choice rules have been examined. 
    As a compromise between equality of opportunity and equality of outcome, it  would be valuable to examine choice rules parameterizing their responsibility: For instance, we can consider the solution $F$ defined as for all $X\in \mathcal{P}$, 
    \begin{equation}
    \label{eq_param}
         F(X) = \argmax_{x\in X} ~ W  \qty( {x_1 \over b_1 (X)^p }, {x_2 \over b_2 (X)^p }, \cdots, {x_n \over b_n (X)^p }), 
    \end{equation}
    where $p\in [0,1]$. 
    The parameter $p$ reflects the extent to which individuals are held responsible for. 
    To the best of the author's knowledge, choice rules that parameterize individual responsibility have only been considered by \citet{karos2018generalization}. 
    They examined the class of proportional choice rules including the egalitarian and  KS choice rules. 
     Providing a characterization of the general class as \eqref{eq_param} could be an interesting direction for future research.
\end{itemize}

\appendix

%%%%%%%%%%%%%%%%%%%%%%%%%%%%%%%%%%%%%%%%%%%%%%%%%%%
\section{Appendix: Proofs of the Results}

In proofs of characterization results, we only prove the only-if parts. 

\subsection{Proof of Theorem \ref{thm_rationalordering}}
%%%%%%%%%%%%%%%%%%%%%%%%%%%%%%%%%%%%%%%%%%%%%%%%%%%

% Note that by \textit{intermediate Pareto} and \textit{anonymity}, for all $x\in \mathbb{R}^n \backslash \{ \mathbf{0} \}$, $x\in F(\text{scmp} \{ x \})$. 
% We use it frequently without any special mention. 

Let $F$ be a choice rule that satisfies \textit{intermediate Pareto}, \textit{scale invariance}, \textit{anonymity}, and \textit{contraction independence for EAI}.  
Define the binary relation $R$ over $\mathbb{R}^n_+$ as for all $x,y\in \mathbb{R}^n_+$, 
\begin{equation}
\label{eq_defrelation}
    x R y \iff x\in F(\text{scmp} \{ x, y\}). 
\end{equation}
The symmetric and asymmetric part of $R$ are denoted by $I$ and $P$, respectively. 

\begin{lem}
\label{lem_ordering}
    Let $F$ be a choice rule that satisfies \textit{intermediate Pareto}, \textit{anonymity}, \textit{contraction independence for EAI}. 
    The binary relation $R$  is complete and transitive. 
\end{lem}

\begin{proof}
    Let $x, y, z \in \mathbb{R}^n_+$.
    By \textit{intermediate Pareto}, there exists $\pi \in \Pi$ such that $x^\pi \in F(\text{scmp} \{ x, y\})$ or $y^\pi \in F(\text{scmp} \{ x, y\})$. 
    By \textit{anonymity}, we have $x\in F(\text{scmp} \{ x, y\})$ or $y \in F(\text{scmp} \{ x, y\})$, that is, $xRy$ or $yRx$. 

    We prove that if $xRy$ and $yRz$, then $xRz$. 
    Consider the problem $\text{scmp} \{ x, y, z\}$. 
    As the argument in the last paragraph, at least one of $x$, $y$, and $z$ is in $F( \text{scmp} \{ x, y, z\})$. 
    We verify that $x \in F( \text{scmp} \{ x, y, z\})$. 
    If $y\in F(\text{scmp}  \{ x, y, z\})$, then by \textit{contraction independence for EAI}, we have $F (\text{scmp}  \{ x, y\}) = F(\text{scmp}  \{ x, y, z\}) \cap \text{scmp}  \{ x, y\}$. 
    By $xRy$, we have $x\in F( \text{scmp} \{ x, y\} )$, which implies $x \in F(\text{scmp}  \{ x, y, z\})$. 
    If $z\in F( \text{scmp} \{ x, y, z\})$, then by \textit{contraction independence for EAI}, we have $F (\text{scmp}  \{ y, z\}) = F(\text{scmp}  \{ x, y, z\}) \cap \text{scmp}  \{ y, z\}$.  
    By $yRz$, we have $y\in F (\text{scmp} \{ y, z\} )$, which implies $y \in F(\text{scmp}  \{ x, y, z\})$. 
    By the result of  the case where $y\in F(\text{scmp}  \{ x, y, z\})$, we have $x \in F(\text{scmp}  \{ x, y, z\})$. 
    Then by \textit{contraction independence for EAI}, $x\in F(\text{scmp}  \{ x, y, z\}) \cap \text{scmp}  \{ x, z\} = F(\text{scmp}  \{ x,  z\})$, that is, $xRz$. 
\end{proof}

By \textit{anonymity} and \textit{scale invariance}, we obtain the following lemma.

\begin{lem}
    Let $F$ be a choice rule that satisfies \textit{intermediate Pareto}, \textit{scale invariance}, \textit{anonymity}, \textit{contraction independence for EAI}. 
    The binary relation $R$ is an weakly monotone,  symmetric and homogeneous ordering. 
\end{lem}

\begin{proof}
    First, we prove that $R$ is weakly monotone. 
    Let $x, y\in \mathbb{R}^n_+$. Suppose that $x\gg y$. By \textit{intermediate Pareto} and \textit{anonymity}, $x\in F(\text{scmp} \{ x, y \})$ and $y\notin F(\text{scmp} \{ x, y \})$. Therefore, we have $x P y$. 
    On the other hand, if $x\geq y$,
    then \textit{intermediate Pareto} and \textit{anonymity} imply that  $x\in F(\text{scmp} \{ x, y \})$. Therefore, we have $x R y$. 

    Then, we prove that $R$ is symmetric.
    Let $x\in  \mathbb{R}^n_+\backslash\{\mathbf{0}\}$ and $\pi \in \Pi$. 
    Since $\text{scmp} \{ x, x^\pi \} = \text{scmp} \{ x \}$, \text{anonymity} implies that  $x, x^\pi\in F(\text{scmp} \{ x, x^\pi \})$. 
    Therefore, we have $x I x^\pi$. 

    Finally, we prove that $R$ is homogeneous, that is, for all $x, y\in  \mathbb{R}^n_+\backslash\{\mathbf{0}\}$ and $\alpha\in \mathbb{R}_{++}$, $x R y \iff \alpha  x R \alpha  y$. Let $x, y\in  \mathbb{R}^n_+\backslash\{\mathbf{0}\}$ and $\alpha \in \mathbb{R}_{++}$ and assume $xRy$, that is, $x\in F(\text{scmp} \{ x, y \})$. By \textit{scale invariance}, this is equivalent to $\alpha x\in F( \alpha \, \text{scmp} \{ x, y \}) = F( \text{scmp} \{ \alpha x,  \alpha y \})$. 
    Therefore, $x Ry$ is equivalent to $\alpha x R \alpha  y$. 
\end{proof}

Then, we prove that for all problems in $\mathcal{P}^e$, $F$ can be represented using the binary relation $R$. 

\begin{lem}
\label{lem_equal_rational}
    Let $F$ be a choice rule that satisfies \textit{intermediate Pareto}, \textit{scale invariance}, \textit{anonymity}, \textit{contraction independence for EAI}. 
    For all $X\in \mathcal{P}^e$,  
    \begin{equation}
        F(X) = \{ x\in X \mid \text{$\not\exists y\in X$ s.t. $y P x$} \}. 
    \end{equation}
\end{lem}

\begin{proof}
    Let $X\in  \mathcal{P}^e$ and $M(X) = \{ x\in X \mid \text{$\not\exists y\in X$ s.t. $y P x$} \}$. 
    First, we prove that $F(X) \subset M(X)$. 
    If $F(X) \not\subset M(X)$, then there exist $x^\ast \in F(X)$ and $z^\ast \in X$ such that $z^\ast P x^\ast$. 
    Since $R$ is transitive (Lemma \ref{lem_ordering}), we can set $z^\ast \in M(X)$. 
    We claim that  $x^\ast \in F(\text{scmp} \, X)$.
    Indeed, if $x^\ast \notin F(\text{scmp} \, X)$, then   by \textit{anonymity}, for some $y\in X$ such that $y\in F(\text{scmp} \, X)$, which means $F(\text{scmp} \, X) \cap  X \neq \emptyset$. 
    By \textit{contraction independence for EAI}, $x^\ast \notin F(\text{scmp} \, X)\cap X = F(X)$, a contradiction. 
    
    Since $\text{scmp} \{ z^\ast, x^\ast \} \subset \text{scmp} \, X$ and $F(\text{scmp} \, X) \cap \text{scmp} \{ z^\ast, x^\ast \} \neq \emptyset$, by \textit{contraction independence for EAI}, 
    \begin{equation*}
        x^\ast \in F(\text{scmp} \, X) \cap \text{scmp} \{ z^\ast, x^\ast \}  =  F(\text{scmp} \{ z^\ast, x^\ast \}), 
    \end{equation*}
    which is a contradiction to $z^\ast P x^\ast$. 

    \vspace{2mm}
    Next, we prove $M(X) \subset F(X)$ for all $X \in \mathcal{P}^e$. 
    If $M(X) \not\subset F(X)$, then there exists $z^{\ast\ast} \in M(X)$ but $z^{\ast\ast} \notin F(X)$. 
    Pick $x^{\ast\ast} \in F(X)$.
    By $F(X) \subset M(X)$, $x^{\ast\ast} \in M(X)$. 
    In the same way as the last paragraph, we can prove $x^{\ast\ast} \in F(\text{scmp} \, X)$. 
    By \textit{contraction independence for EAI},  $z^{\ast\ast} \notin F( X) = F(\text{scmp} \, X) \cap  X$, i.e., $z^{\ast\ast}\notin F(\text{scmp} \, X) $.
    Again by \textit{contraction independence for EAI},  
    \begin{equation*}
        z^{\ast\ast} \notin F(\text{scmp} \, X)  \cap  \text{scmp} \{ z^{\ast\ast}, x^{\ast\ast} \} = F(\text{scmp} \{ z^{\ast\ast}, x^{\ast\ast} \}), 
    \end{equation*}
    that is, $x^{\ast \ast} P z^{\ast \ast}$. 
    This is a contradiction to $z^{\ast\ast} \in M(X)$.
\end{proof}

\begin{proof}[\bf Proof of Theorem \ref{thm_rationalordering}]
    Take $X \in \mathcal{P}$ be an arbitrary problem. 
    By \textit{scale invariance}, 
    \begin{equation}
    \label{eq_SInomalized_rational}
        F(X) = b(X) F\qty( \qty( {1 \over b_1 (X)},{1 \over b_2 (X)}, \cdots, {1 \over b_n (X)}   ) X ).
    \end{equation}
    Let 
    \begin{equation*}
        X^\ast = \qty( {1 \over b_1 (X)},{1 \over b_2 (X)}, \cdots, {1 \over b_n (X)}  ) X \in \mathcal{P}^e. 
    \end{equation*}
    By Lemma \ref{lem_equal_rational},
    \begin{align*}
        F(X^\ast) &= \{ x\in X^\ast \mid \text{$\not\exists y\in X^\ast$ s.t. $y P x$} \}. 
    \end{align*} 
    By \eqref{eq_SInomalized_rational}, we have 
     \begin{align*}
        F(X) & = b(X) \{ x\in X^\ast \mid \text{$\not\exists y\in X^\ast$ s.t. $y P x$} \}  \\
        & = \qty{ x\in X ~ \bigg| ~ \text{$\not\exists y\in X$ s.t. $\qty( {y_1\over b_1(X)}, \cdots, {y_n\over b_n(X)} )  P \qty( {x_1\over b_1(X)}, \cdots, {x_n\over b_n(X)} ) $} }. 
    \end{align*}  
    that is, $F$ is rationalized by the normalized welfare ordering $R$ defined as \eqref{eq_defrelation}. 
\end{proof}

\subsection{Proof of Theorem \ref{thm_rational}}

Let $F$ be a choice rule that satisfies \textit{intermediate Pareto}, \textit{scale invariance}, \textit{anonymity}, \textit{contraction independence for EAI}, and \textit{continuity}.

We start with simple observations about the problems of the form $\text{scmp} \{ x, \gamma \mathbf{1} \}$, where $x\in \mathbb{R}^n_{+}\backslash \{ \mathbf{0} \} $ and $\gamma \in \mathbb{R}_+$. 

\begin{lem}
\label{lem_thre1}
 Let $F$ be a choice rule that satisfies \textit{intermediate Pareto}, \textit{anonymity}, \textit{contraction independence for EAI}, and \textit{continuity}. For all $x\in \mathbb{R}^n_+\backslash\{\mathbf{0}\}$, there exists $\gamma \in \mathbb{R}_{+}$ such that $ \{x, \gamma \mathbf{1} \} \subset F(\text{scmp} \{ x, \gamma \mathbf{1} \})$. 
\end{lem}

\begin{proof}
For all $x\in \mathbb{R}^n_+\backslash\{\mathbf{0}\}$, let $L_x^+ = \{ \alpha \in \mathbb{R}_{+} \mid \alpha \mathbf{1} \in F(\text{scmp}\{ x, \alpha \mathbf{1}  \}) \}$ and $L_x^- = \{ \alpha \in \mathbb{R}_{+} \mid x \in F(\text{scmp}\{ x, \alpha \mathbf{1}  \}) \}$. 
    By \textit{intermediate Pareto} and \textit{anonymity}, $ F(\text{scmp}\{ x, \alpha \mathbf{1}  \}) \cap \{ x, \alpha \mathbf{1}  \} \neq \emptyset$ for all $\alpha \in \mathbb{R}_+$, which implies  $L^+_x \cup L^-_x = \mathbb{R}_{+}$. 
    By \textit{intermediate Pareto}, $\alpha \mathbf{1} \in  F(\text{scmp}\{  \alpha \mathbf{1} \} )  = F(\text{scmp}\{ x, \alpha \mathbf{1} \} ) $ for $\alpha \geq \max_{i\in N} x_i$ and $x \in F(\text{scmp}\{ x \} ) =  F(\text{scmp}\{ x, \alpha \mathbf{1}\} ) $ for $\alpha \leq \min_{i\in N} x_i$. 
    Therefore, $L^+_x \neq \emptyset$ and  $L^-_x \neq \emptyset$. 
    By \textit{continuity}, $L^+_x$ and  $L^-_x$ are closed. 
    Since $\mathbb{R}_+$ is connected, $L^+_x \cap L^-_x  \neq \emptyset$, that is, for some $\gamma \in \mathbb{R}_{+}$, $ \{x, \gamma \mathbf{1} \} \subset F(\text{scmp} \{ x, \gamma \mathbf{1} \})$. 
\end{proof}

\begin{lem}
    \label{lem_thre2}
    Let $F$ be a choice rule that satisfies \textit{intermediate Pareto}, \textit{anonymity}, \textit{contraction independence for EAI},  and \textit{continuity}, 
    and suppose that  $\{ x, \gamma \mathbf{1}\} \subset F(\text{scmp} \{ x, \gamma \mathbf{1}\}  ) $ for $x\in \mathbb{R}^n_{+}\backslash \{\mathbf{0}\}$ and $\gamma \in \mathbb{R}_+$. Then the following statements hold: \\
     (i) For all $\delta \in [0, \gamma) $, $x\in F(\text{scmp} \{ x, \delta \mathbf{1} \})$ and $\delta \mathbf{1} \notin F(\text{scmp} \{ x, \delta \mathbf{1} \}) $.
     \\
     (ii)  For all $\delta > \gamma$, $\delta \mathbf{1} \in F(\text{scmp} \{ x, \delta \mathbf{1} \})$ and $x\notin  F(\text{scmp} \{ x, \delta \mathbf{1} \})$.
\end{lem}

\begin{proof}
    (i) Let $\delta \in  [0, \gamma) $.   
    By \textit{contraction independence for EAI}, 
    \begin{equation*}
        x\in  F(\text{scmp} \{ x, \gamma \mathbf{1} \}) \cap \text{scmp} \{ x, \delta \mathbf{1} \} =
        F(\text{scmp} \{ x, \delta \mathbf{1} \}). 
    \end{equation*}
    If $\delta \mathbf{1} \in F(\text{scmp} \{ x, \delta \mathbf{1} \})$, then we have $\{ x,\delta \mathbf{1} \} \in F(\text{scmp} \{ x, \delta \mathbf{1} \})$. 
    Since $F(\text{scmp} \{ x,\gamma \mathbf{1} \}) \cap  \text{scmp} \{ x,\delta \mathbf{1} \} \neq \emptyset$,  \textit{contraction independence for EAI} implies 
    \begin{equation*}
        \delta\mathbf{1} \in F(\text{scmp} \{ x,\delta \mathbf{1} \})= F(\text{scmp} \{ x,\gamma \mathbf{1} \}) \cap \{ x,\delta \mathbf{1} \}, 
    \end{equation*}
    that is $\delta\mathbf{1} \in  F(\text{scmp} \{ x,\gamma \mathbf{1} \})$. This is a contradiction to \textit{intermediate Pareto}. 
    
\vspace{3mm}
\noindent
    (ii) Let $\delta > \gamma$. 
    Suppose to the contrary that $\delta \mathbf{1} \notin F(\text{scmp} \{ x, \delta \mathbf{1} \})$. 
    By  \textit{intermediate Pareto} and \textit{anonymity}, $x\in F(\text{scmp} \{ x, \delta \mathbf{1} \})$, which implies $F(\text{scmp} \{ x, \delta \mathbf{1} \}) \cap \text{scmp} \{ x, \gamma \mathbf{1} \} \neq \emptyset$. By \textit{intermediate Pareto} and  \textit{contraction independence for EAI},
    \begin{equation*}
        \gamma \mathbf{1} \notin F(\text{scmp} \{x, \delta \mathbf{1}\} ) \cap \text{scmp} \{ x, \gamma \mathbf{1} \}  = F(\text{scmp} \{ x, \gamma \mathbf{1} \}), 
    \end{equation*}
    a contradiction. 
    If  $x\in F(\text{scmp} \{ x, \delta \mathbf{1} \})$, then by (i) and the result of the last paragraph, $\gamma \mathbf{1} \notin F(\text{scmp} \{ x, \gamma \mathbf{1} \}) $. This is a contradiction to $\{ x, \gamma \mathbf{1}\} \subset F(\text{scmp} \{ x, \gamma \mathbf{1}\}  ) $. 
\end{proof}

Now we define the function $W:\mathbb{R}^n_+  \rightarrow \mathbb{R}_+$ as for all $x\in \mathbb{R}^n_+ $, 
\begin{equation}
\label{eq_pridefW}
    W(x) = \inf \{ \alpha \in \mathbb{R}_+ \mid \alpha > 0 ~~\text{and} ~~\alpha \mathbf{1} \in F( \text{scmp} \{ x, \alpha \mathbf{1} \}  ) \}. 
\end{equation}
By Lemma \ref{lem_thre1} and \ref{lem_thre2}, for all $x\in \mathbb{R}^n_+ \backslash \{ \mathbf{0} \}$,
\begin{equation}
\label{eq_Wpro1}
     \{ x, W(x)\mathbf{1}\} \subset F(\text{scmp} \{ x,  W(x)\mathbf{1}\})
\end{equation}
and $W(\mathbf{0}) = 0$. 
Also, note that by \eqref{eq_Wpro1} and Lemma \ref{lem_thre2},  for all $x\in \mathbb{R}^n_+ \backslash \{ \mathbf{0} \}$ and $\alpha \in \mathbb{R}_+$,
\begin{equation}
\label{eq_Wproperty}
    % F(\text{scmp} \{ x, \alpha \mathbf{1} \}) 
    % \subset 
    \left\{
    \begin{array}{ll}
    \alpha \mathbf{1} \in F(\text{scmp} \{ x, \alpha \mathbf{1} \}), & \alpha  > W(x) \\
    \{ x, \alpha \mathbf{1} \} \subset F(\text{scmp} \{ x, \alpha \mathbf{1} \}), & \alpha  = W(x) \\
     x \in F(\text{scmp} \{ x, \alpha \mathbf{1} \}), & \alpha  < W(x). 
    \end{array}
    \right.
\end{equation}

% We show that the function $W$ is weakly monotone.
% Let $x, y\in \mathbb{R}^n_+$ such that $x_i \gg y_i $. By \textit{intermediate Pareto}, $x \in F(\text{scmp} \{ x, y \})$ and $y \notin F(\text{scmp} \{ x, y \})$. By Lemma \ref{lem_representedW}, $W(x) > W(y)$. 

\begin{proof}[\bf Proof of Theorem \ref{thm_rational}]
    By Theorem \ref{thm_rationalordering}, there exists a weakly monotone, symmetric and homogeneous weak order $R$ over $\mathbb{R}^n_+$ such that $F$ is  is rationalized by the normalized welfare ordering $R$.  
    
    We claim that for all $x,y \in \mathbb{R}^n_+$, $xRy \iff W(x) \geq W(y)$. 
    To prove this, let $x,y \in \mathbb{R}^n_+$ with $xRy$.  By construction of $W$, $x I\, W(x) \mathbf{1}$ and $y I \,W(y) \mathbf{1}$. By transitivity of $R$, $xRy$ is equivalent to $W(x) \mathbf{1}\, R\,  W(y) \mathbf{1}$. Since $R$ is weakly monotone, this is equivalent to $W(x) \geq W(y)$. 
    
    Therefore, for all $X\in \mathcal{P}$, 
    \begin{align*}
            F(X)  &= 
            \qty{ x\in X ~ \bigg| ~ \text{$\not\exists y\in X$ s.t. $\qty( {y_1\over b_1(X)}, \cdots, {y_n\over b_n(X)} )  P \qty( {x_1\over b_1(X)}, \cdots, {x_n\over b_n(X)} ) $} }\\
            &= 
            \argmax_{x\in X} ~ W \qty(  {x_1\over b_1(X)}, {x_2\over b_2(X)}, \cdots, {x_n\over b_n(X)} ). 
    \end{align*}  
    Since $R$ is weakly monotone, symmetric, and homogeneous, it is straightforward to prove that $W$ is also weakly monotone, symmetric, and homogeneous. 

    Finally, we prove that $W$ is continuous. 
    Suppose to the contrary that there exists $x\in \mathbb{R}^n_+ $ such that $W$ is discontinuous at $x$. 
    Then there exist $\varepsilon > 0$ and a sequence $\{ x^n \}_{n\in \mathbb{N}} \subset \mathbb{R}^n_+\backslash \{\mathbf{0} \}$ such that  $x^n\rightarrow x$ as $n \rightarrow +\infty$ but $|W(x^n) - W(x)| > \varepsilon$ for all $n\in \mathbb{N}$. 
    Without loss of generality, we assume $W(x^n) > W(x) + \varepsilon $ for all $n\in \mathbb{N}$.
    By \eqref{eq_Wproperty}, for all $n \in \mathbb{N}$, $x^n\in F(\text{scmp} \{ x^n , (W(x) + \varepsilon)  \mathbf{1}\}) $.
    By \textit{continuity}, we have $x\in F(\text{scmp} \{ x , (W(x) + \varepsilon)  \mathbf{1}\}) $. 
    If $x\neq \mathbf{0}$, then this is a contradiction to \eqref{eq_Wproperty}. 
    If  $x = \mathbf{0}$, we have $\mathbf{0} \in F(\text{scmp} \{ \varepsilon \mathbf{1}\} )$, which is a contradiction to \textit{intermediate Pareto}. 
\end{proof}

%%%%%%%%%%%%%%%%%%%%%%%%%%%%%%%%%%%%%%%%%%%%%%%%
\subsection{Proof of Theorem \ref{thm_main}}
%%%%%%%%%%%%%%%%%%%%%%%%%%%%%%%%%%%%%%%%%%%%%%%%

We prove Theorem \ref{thm_main} using Theorem \ref{thm_rational} and the arguments in its proof, and then verify the independence of the axioms. 

Let $F$ be a choice rule that satisfies \textit{intermediate Pareto}, \textit{scale invariance},  \textit{anonymity}, \textit{contraction independence for EAI}, \textit{continuity}, \textit{equal addition independence for EAI}, and \textit{compromisability for EAI}.

\begin{lem}
\label{lem_propW1}
    Then, the function $W$ defined in \eqref{eq_pridefW} satisfies the following property: For all $x\in \mathbb{R}^n_+$ and $\beta \in \mathbb{R}_{++}$,
    \begin{equation}
    \label{eq_commonadd}
        W(x+\beta \mathbf{1}) = W(x) + \beta.
    \end{equation}
    Furthermore, it can be uniquely extended to $\widetilde{W}$ defined on $\mathbb{R}^n$ such that $\widetilde{W}$ is a homogeneous, weakly monotone, symmetric, continuous function and satisfies the counterpart of \eqref{eq_commonadd}: For all $x\in \mathbb{R}^n$ and $\beta \in \mathbb{R}_{++}$,
    \begin{equation}
    \label{eq_commonadd2}
        \widetilde{W} (x+\beta \mathbf{1}) = \widetilde{W} (x) + \beta.
    \end{equation}
\end{lem}

\begin{proof}
    First, we prove the function $W$  satisfies \eqref{eq_commonadd}. 
    Let $x\in \mathbb{R}^n_+ \backslash \{\mathbf{0} \}$ and $\beta \in \mathbb{R}_{++}$. 
    Since $\{ x, W(x) \mathbf{1}\} \subset F(\text{scmp} \{ x , W(x)  \mathbf{1}\})$, by \textit{equal addition independence for EAI}, $\{ x+  \beta\mathbf{1} , (W(x)+\beta) \mathbf{1}\} \subset F(\text{scmp} \{ x+  \beta\mathbf{1} , (W(x)+  \beta)\mathbf{1}\})$, that is, $W(x+\beta \mathbf{1}) = W(x) + \beta$. 
    
    Since $W(\mathbf{0}) = 0$, we have $W(\mathbf{0}+ \beta \mathbf{1}) = W(\beta \mathbf{1}) =  \beta =  W(\mathbf{0}) + \beta$. 
    Therefore, for all $x\in \mathbb{R}^n_+$,  and $\beta \in \mathbb{R}_{++}$, $ W(x+\beta \mathbf{1}) = W(x) + \beta$. 

    Then, we prove the second part. 
    Given $W$, define $\widetilde{W}: \mathbb{R}^n \rightarrow \mathbb{R}$ as for all $x\in \mathbb{R}^n$, 
    $\widetilde{W} (x) = W(x + \alpha \mathbf{1}) -  \alpha$ for some 
    $\alpha$ such that $x + \alpha \mathbf{1} \in \mathbb{R}^n_{+}$. 
    We verify that this function is well-defined. 
    Let $\alpha, \beta \in \mathbb{R}_{++}$ be such that $x + \alpha \mathbf{1}, x + \beta \mathbf{1} \in \mathbb{R}^n_{+} $ and $\alpha > \beta$. By \eqref{eq_commonadd}, $ W(x + \alpha \mathbf{1}) -  \alpha =  W(x + \beta \mathbf{1}) +(\alpha - \beta) -\alpha =  W(x + \beta \mathbf{1}) - \beta $. 
    
     It is straightforward to prove that $\widetilde{W}$ is a homogeneous, weakly monotone, symmetric, continuous function and satisfies \eqref{eq_commonadd2}. 
     Finally, we prove the uniqueness. 
     Suppose to the contrary that there is two different function $\widetilde{W}$ and $\overline{W}$ satisfying all of the property in the statement. Then there exists $x\in \mathbb{R}^n$ such that $\widetilde{W} (x) \neq \overline{W} (x)$. By \eqref{eq_commonadd2},  there exists $y\in \mathbb{R}^n_{+}$ such that $\widetilde{W} (y) \neq \overline{W} (y)$. However, since both of $\widetilde{W}$ and $\overline{W}$ are extensions of $W$, this is a contradiction. 
\end{proof}

Next, we show that the upper contour set of each vector is convex. For each $x\in \mathbb{R}^n$, let $U^+ (x) = \{  y\in \mathbb{R}^n_+  \mid W (y) \geq W (x) \}$ and $\widetilde{U}^+ (x) = \{  y\in \mathbb{R}^n \mid \widetilde{W} (y) \geq \widetilde{W} (x) \}$.

\begin{lem}
\label{lem_propW2}
    Let $F$ be a choice rule that satisfies \textit{intermediate Pareto}, \textit{scale invariance},  \textit{anonymity}, \textit{contraction independence for EAI}, \textit{continuity}, \textit{equal addition independence for EAI}, and \textit{compromisability for EAI}. 
    Then $\widetilde{W}$ defined in Lemma \ref{lem_propW1} satisfies the following property: For each $x \in \mathbb{R}^n$, the  set $\widetilde{U}^+ (x)$ is convex. 
\end{lem}

\begin{proof}
Since $\widetilde{W}$  satisfies \eqref{eq_commonadd2}, it is sufficient to consider vectors in $\mathbb{R}^n_{+} $, i.e., to prove that for each $z \in \mathbb{R}^n_{++} $, the  set $U^+ (z)$ is convex. 

Suppose to the contrary that there exists $x, y \in \mathbb{R}^n_{+}$ and $\alpha \in (0,1)$ such that $x, y \in U^+ (z)$ but $\alpha x + (1- \alpha ) y \notin  U^+ (z)$. 
Without loss of generality, we assume $W(x) = W(y) = W(z)$. 
Indeed, if $W(x) > W(z)$, then by $W(z) > W(\alpha x + (1- \alpha ) y)$ and the continuity of $W$ (cf. Theorem \ref{thm_rational}), the intermediate value theorem implies there exists $x^\ast$ such that $W(x^\ast) = W(z)$ and $x^\ast$ is in the line segment joining $x$ and $\alpha x + (1- \alpha ) y$. Similarly, if  $W(y) > W(z) $, then we can take $y^\ast$ such that $W(y^\ast) = W(z)$ and $y^\ast$ is in the line segment joining $y$ and $\alpha x + (1- \alpha ) y$.

We claim that neither $x\geq y$ nor $y\geq x$ holds. 
If $x\geq y$, then $W(y) = W(z) > W (\alpha x + (1- \alpha ) y ) $. 
However, this is a contradiction to $\alpha x + (1- \alpha ) y \geq y$ and the fact that $W$ is weakly monotone and continuous (cf. Theorem \ref{thm_rational}). 

Let $X = \text{scmp} \{ x, y, \alpha x + (1- \alpha ) y \}$. 
By Theorem \ref{thm_rational}  and $W(x) = W(y) > W( \alpha x + (1- \alpha ) y )$, $\{x, y \} \subset F(X)$. By \textit{compromisability for EAI}, there exists $s \geq  \alpha x + (1- \alpha ) y$ such that $s \in F(X)$. Since neither $x\geq y$ nor $y\geq x$ holds, there is no $s$ with $s > \alpha x + (1- \alpha ) y$. Therefore, $\alpha x + (1- \alpha ) y \in F(X)$, which is a contradiction to Theorem \ref{thm_rational}. 
\end{proof}

\begin{lem}
\label{lem_normalizedmaximin}
    Let $F$ be a choice rule that satisfies \textit{intermediate Pareto}, \textit{scale invariance},  \textit{anonymity}, \textit{contraction independence for EAI}, \textit{continuity}, \textit{equal addition independence for EAI}, and \textit{compromisability for EAI}. 
    Let $\widetilde{W}$ be the function defined in Lemma \ref{lem_propW1}. 
    Then there exists a nonempty symmetric closed  convex set  $\mathcal{W} \subset \Delta$ such that for all $x\in \mathbb{R}^n$
    \begin{equation*}
        \widetilde{W}(x) = \min_{w \in \mathcal{W}} \sum_{i\in N} w_i x_i. 
    \end{equation*}
\end{lem}

\begin{proof}
We prove this lemma step-by-step. 

\noindent
\textbf{Step 1. There exists a nonempty closed convex set  $\mathcal{W} \subset \Delta$ such that
\begin{equation}
\label{eq_tildeW_maximin}
    \widetilde{W} (x) = \min_{w\in \mathcal{W}} \sum_{i\in N} w_i x_i
\end{equation}
for all $x\in \mathbb{R}^n$.}

Consider the set $\widetilde{U}^+ (\mathbf{0})$. 
By Lemma \ref{lem_propW2}, $\widetilde{U}^+ (\mathbf{0})$ is convex. 
By continuity of $\widetilde{W}$ (cf. Lemma \ref{lem_propW1}), $\widetilde{U}^+ (\mathbf{0})$ is  closed.  
Since homogeneity of $\widetilde{W}$ (cf. Lemma \ref{lem_propW1}) implies that  $\alpha x \in \widetilde{U}^+ (\mathbf{0})$ for all $x \in \widetilde{U}^+ (\mathbf{0})$ and  $\alpha \in \mathbb{R}_{++}$,
the set $\widetilde{U}^+ (\mathbf{0})$ is a nonempty closed convex cone. 
By applying the supporting hyperplane theorem to $\widetilde{U}^+ (\mathbf{0})$ and $\mathbf{0}$, there exists $w\in \Delta$ such that for all $x\in \widetilde{U}^+ (\mathbf{0})$, 
\begin{equation}
\label{eq_support}
    \sum_{i\in N} w_i x_i \geq 0. 
\end{equation}

Let $\mathcal{W}\subset \Delta$ be a set of vectors $w$ satisfying \eqref{eq_support} for all $x \in \widetilde{U}^+ (\mathbf{0})$.  
The set $\mathcal{W}$ is convex. Indeed, if $w, w'\in \mathcal{W}$ and $\alpha\in[0,1]$, then for all $x \in \widetilde{U}^+ (\mathbf{0})$, 
\begin{equation*}
     \sum_{i\in N} w_i x_i \geq 0 ~~~\text{and}~~~  \sum_{i\in N} w'_i x_i \geq 0, 
\end{equation*}
which implies 
\begin{equation*}
     \sum_{i\in N} (\alpha w_i + (1 - \alpha) w'_i) x_i \geq 0. 
\end{equation*}

We claim that $\mathcal{W}$ is a closed set. 
Let $\{ w^k \}_{k \in \mathbb{N}} \subset \mathcal{W}$ be a sequence that converges to $w$. By the definition of $\mathcal{W}$, for all  $k\in \mathbb{N}$ and $x\in \widetilde{U}^+ (\mathbf{0})$, 
\begin{equation*}
     \sum_{i\in N} w^k_i x_i \geq 0. 
\end{equation*}
Since $\{ w^k \}_{k \in \mathbb{N}}$ converges to $w$, we have 
     $\sum_{i\in N} w_i x_i \geq 0$ for all $x\in \widetilde{U}^+ (\mathbf{0})$, 
that is, $w\in \mathcal{W}$. 

We claim that for all $x\in \mathbb{R}^n$,  $\widetilde{W}(x)\geq 0$ if and only if $\min_{w\in \mathcal{W}} \sum_{i\in N} w_i x_i \geq 0$. 
Let $x\in \mathbb{R}^n$ with $\widetilde{W} (x)\geq 0$. Then for all $w\in \mathcal{W}$, $\sum_{i\in N} w_i x_i \geq 0$, that is, $\min_{w\in \mathcal{W}} \sum_{i\in N} w_i x_i \geq 0$.
For the converse, let $x\in \mathbb{R}^n$ be such that $\min_{w\in \mathcal{W}} \sum_{i\in N} w_i x_i \geq 0$ but $\widetilde{W}(x) <  0$ (i.e., $x \notin \widetilde{U}^+ (\mathbf{0})$). Since $\widetilde{U}^+ (\mathbf{0})$ is a nonempty closed convex cone, there exists $w' \in \mathcal{W}$ such that $\sum_{i\in N} w'_i x_i < 0$, which is a contradiction. 

Assume that $\widetilde{W}(x) = 0$ and $\min_{w\in \mathcal{W}} \sum_{i\in N} w_i x_i > 0$. 
Let $\varepsilon \in \mathbb{R}_{++}$ with $0 < \varepsilon < \min_{w\in \mathcal{W}}  \sum_{i\in N} w_i x_i$. 
Then we have  $\min_{w\in \mathcal{W}} \sum_{i\in N} w_i (x_i - \varepsilon)  > 0$. By the result of the last paragraph, we have $\widetilde{W}(x - \varepsilon \mathbf{1}) \geq 0$. 
Since $\widetilde{W}$ is weakly monotone, we have $0 = \widetilde{W}(x) > \widetilde{W}(x - \varepsilon \mathbf{1}) \geq 0$, a contradiction. 
Therefore, if  $\widetilde{W}(x) = 0$, then $\min_{w\in \mathcal{W}} \sum_{i\in N} w_i x_i = 0$. 

Finally, we prove that $\widetilde{W}$ can be written as \eqref{eq_tildeW_maximin}. For each $x\in \mathbb{R}^n$, by \eqref{eq_commonadd2}, we can take $x^\ast \in \mathbb{R}^n$ such that $\widetilde{W} (x^\ast) = 0$ and $x = x^\ast + \widetilde{W} (x) \mathbf{1}$. 
Therefore, by the result of the last paragraph, 
\begin{align*}
    \widetilde{W} (x) &= \widetilde{W} (x^\ast +   \widetilde{W}(x) \mathbf{1}) \\
    &= \widetilde{W} (x^\ast) + \widetilde{W} (\widetilde {W} (x) \mathbf{1}) \\
    &= \qty( \min_{w\in \mathcal{W}} \sum_{i\in N} w_i x^\ast_i ) + \widetilde {W} (x) \\
    &= \min_{w\in \mathcal{W}} \sum_{i\in N} w_i (x^\ast_i + \widetilde {W} (x)\mathbf{1} )\\
    &= \min_{w\in \mathcal{W}}  \sum_{i\in N} w_i x_i.
\end{align*}

\noindent
\textbf{Step 2. The set $\mathcal{W}$ is symmetric.}

Suppose to the contrary that there exist $w\in \mathcal{W}$ and $\pi \in \Pi$ such that $w^\pi \notin \mathcal{W}$. 
Since $\mathcal{W}$ is a closed convex set, by the separating hyperplane theorem, there exist $x\in \mathbb{R}^n$ and $\gamma\in \mathbb{R}$ such that for all $v\in \mathcal{W}$
\begin{equation}
\label{eq_separation}
    \sum_{i\in N} w^{\pi}_i x_i < \gamma < \sum_{i\in N} v_i x_i. 
\end{equation}
Note that for all $(\alpha , \beta) \in \mathbb{R}_{++}\times \mathbb{R}$, if a pair $(x, \gamma)\in \mathbb{R}^n\times \mathbb{R}$ satisfies the above inequalities, then the pair $(\alpha x + \beta \mathbf{1}, \alpha \gamma + \beta)$ also satisfies the inequalities. Thus, we can assume that $x_i, \gamma\in (0,1)$ for all $i\in N$.

By \eqref{eq_separation}, for all $v\in \mathcal{W}$, $\sum_{i\in N} v_i \gamma= \gamma < \sum_{i\in N} v_i x_i$. 
By Step 1, we have $\widetilde {W}(x) > \widetilde {W}(\gamma \mathbf{1})$. 
% Therefore, we have $F(\text{scmp} \{ x, \gamma \mathbf{1} \}) \subset \text{scmp} \{ x\}$. 
% By \textit{intermediate Pareto} and \textit{anonymity}, $x\in F(\text{scmp} \{ x, \gamma \mathbf{1} \}) $.  
Let $y\in\mathbb{R}^n$ with $y^\pi = x$. 
Again by \eqref{eq_separation}, 
\begin{equation*}
    \sum_{i\in N} w_i y_i = \sum_{i\in N} w^{\pi}_i y^\pi_i  = \sum_{i\in N} w^{\pi}_i x_i < \gamma =  \sum_{i\in N} w_i \gamma.
\end{equation*}
By Step 1 and  $w\in \mathcal{W}$, $\widetilde {W}(y) < \widetilde {W}(\gamma \mathbf{1})$, which is a contradiction to the fact that $\widetilde {W}$ is symmetric (cf. Lemma \ref{lem_propW1}). 
\end{proof}

\begin{proof}[\bf Proof of Theorem \ref{thm_main}]
    Take $X \in \mathcal{P}$ be an arbitrary problem. By \textit{scale invariance}, 
    \begin{equation}
    \label{eq_SInomalized}
        F(X) = b(X) F\qty( \qty( {1 \over b_1 (X)},{1 \over b_2 (X)}, \cdots, {1 \over b_n (X)}   ) X ).
    \end{equation}
    Let 
    \begin{equation*}
        X^\ast = \qty( {1 \over b_1 (X)},{1 \over b_2 (X)}, \cdots, {1 \over b_n (X)}  ) X \in \mathcal{P}^e. 
    \end{equation*}
    By Theorem \ref{thm_rational} and Lemma \ref{lem_normalizedmaximin}, there exists a nonempty symmetric  closed  convex set $\mathcal{W}$ such that
    \begin{align*}
        F(X^\ast) &= \argmax_{x^\ast \in X^\ast}\min_{w \in \mathcal{W}} \sum_{i\in N} w_i x^\ast_i. 
    \end{align*} 
    By \eqref{eq_SInomalized}, we have 
     \begin{align*}
        F(X)  = b(X) \qty( \argmax_{x^\ast \in X^\ast}\min_{w \in \mathcal{W}} \sum_{i\in N} w_i x^\ast_i ) = \argmax_{x\in X}\min_{w \in \mathcal{W}} \sum_{i\in N} w_i {x_i\over b_i(X)}. 
    \end{align*} 
\end{proof}

We verify the independence of the axioms in Theorem \ref{thm_main}.

\begin{example}
    The KS choice rule satisfies all the axioms except for \textit{intermediate Pareto}. 
\end{example}

\begin{example}
    Let $F$ be the choice rule such that for all $X\in \mathcal{P}$, 
    \begin{equation*}
        F(X) = \argmax_{x\in X} \min_{i\in N} x_i.
    \end{equation*}
    This choice rule satisfies all the axioms except for \textit{scale invariance}. 
\end{example}

\begin{example}
    Let $F$ be the choice rule such that for all $X\in \mathcal{P}$, 
    \begin{equation*}
        F(X) = \argmax_{x\in X} x_1.
    \end{equation*}
    This choice rule satisfies all the axioms except for \textit{anonymity}. 
\end{example}

\begin{example}
    Let $F$ be the choice rule that assigns to each $X\in \mathcal{P}$ the set of all weakly Pareto optimal utility vectors, that is, 
    \begin{equation*}
        F(X) = \{ x \in X \mid \text{$\not\exists y\in X$ s.t. $y\gg x$ }\}.
    \end{equation*}
    This choice rule satisfies  all the axioms except for \textit{contraction independence for EAI}. 
\end{example}

\begin{example}
   Define the binary relation $>^\text{lex}$ over $\mathbb{R}^n$ as for all $x, y\in \mathbb{R}^n$, 
   \begin{align*}
       x >^\text{lex} y \iff \Big[ \text{$ \exists j \leq n$ \, s.t. \,} x_{(i)} = y_{(i)} ~ \text{for all $i<  j$ and }  x_{(j)} > y_{(j)}\Big].
   \end{align*}
   % \begin{align*}
   %     x \geq^\text{lex} y \iff &\Big[ \text{there exists $j\geq n$ such that } x_{(i)} = y_{(i)} ~ \text{for all $i\leq j$ and }  x_{(i)} > y_{(i)}\Big], \\
   %     &\text{or} \Big[x_{(j)} = y_{(j)} ~ \text{for all $i\leq n$}\Big].
   % \end{align*}
   % The asymmetric part of $\geq^\text{lex}$ is denoted by $>^\text{lex}$. 
   Let $F$ be the choice rule that assigns to $X\in \mathcal{P}$ the set 
   \begin{equation*}
       \qty{ x\in X ~ \Bigg| \not\exists y \in X ~~ \text{s.t.}~~ \qty( {y_1\over b_1(X)} , \cdots,  {y_n\over b_n(X)} ) >^\text{lex} \qty( {x_1\over b_1(X)} , \cdots,  {x_n\over b_n(X)} ) }. 
   \end{equation*}
   This choice rule  satisfies  all the axioms except for \textit{continuity}. 
\end{example}

\begin{example}
   The \citet{nash1950bargaining} choice rule, which assigns to each $X\in \mathcal{P}$ the maximizers of $\prod_{i\in N} x_i$ over $x\in X$, satisfies all the axioms except for the  \textit{equal addition independence for EAI}. 
\end{example}

\begin{example}
     Let $F$ be the choice rule such that for all $X\in \mathcal{P}$, 
    \begin{equation*}
        F(X) = \argmax_{x\in X} \max_{i\in N} {x_i\over b_i (X)}.
    \end{equation*}
    This choice rule satisfies all the axioms except for \textit{compromisability for EAI}. 
\end{example}

%%%%%%%%%%%%%%%%%%%%%%%%%%%%%%%%%%%%%%%%%%%%%%%%
\subsection{Proof of Proposition \ref{prop_relmaximin}}
%%%%%%%%%%%%%%%%%%%%%%%%%%%%%%%%%%%%%%%%%%%%%%%%

\begin{proof}[\bf Proof of Proposition \ref{prop_relmaximin}]
     Let $F$ be a choice rule that satisfies \textit{intermediate Pareto}, \textit{scale invariance}, \textit{anonymity}, \textit{contraction independence for EAI}, \textit{continuity}, and \textit{Hammond equity for EAI}.  
     By Theorem \ref{thm_rational}, 
    $F$ is rationalized by a normalized welfare function $W:\mathbb{R}^n_+ \rightarrow \mathbb{R}_+$ such that $W$ is weakly monotone and continuous. 

    Let $x, y \in \mathbb{R}^n_{+}$ be such that 
    \begin{align*}
        \begin{array}{cc}
    x_i < y_i < y_j < x_j, & \text{for some }i, j \in N, \\
   x_k = y_k, & \text{for all }k\in N \backslash\{ i, j \}. 
    \end{array}
    \end{align*}
    Consider the problem $\text{scmp} \{ x, y \}$. By \textit{intermediate Pareto} and \textit{anonymity}, $x\in F(\text{scmp} \{ x, y \})$ or $y\in F(\text{scmp} \{ x, y \})$ holds. By \textit{Hammond equity for EAI}, we have  $y\in F(\text{scmp} \{ x, y \})$. 
    By Theorem \ref{thm_rational}, we have $W(y) \geq W(x)$. 
    By Theorem in \citet{miyagishima2010characterization}, we have $W(x) = \min_{i\in N} x_i$ for all $x\in \mathbb{R}^n_{+}$.\footnote{\citet{miyagishima2010characterization} examined social welfare orderings over $\mathbb{R}^n$. Although we derived properties of $W$, it is straightforward to rewrite them as properties of orderings. 
    Also, the domain of orderings in the result of \citet{miyagishima2010characterization} can be changed from $\mathbb{R}^n$ to $\mathbb{R}^n_{+}$ without any change of the proof. 
    Hence, we can apply this result. }
\end{proof}

%%%%%%%%%%%%%%%%%%%%%%%%%%%%%%%%%%%%%%%%%%%%%%%%
\subsection{Proof of Proposition \ref{prop_relutil}}
%%%%%%%%%%%%%%%%%%%%%%%%%%%%%%%%%%%%%%%%%%%%%%%%

 In this section, we assume that $n\geq 3$. 

\begin{proof}[\bf Proof of Proposition \ref{prop_relutil}]
    Let $F$ be a choice rule that satisfies \textit{intermediate Pareto}, \textit{scale invariance}, \textit{anonymity}, \textit{contraction independence for EAI}, \textit{continuity}, \textit{equal addition independence for EAI}, and \textit{separability for EAI}. 
    By Theorem \ref{thm_rational} and Lemma \ref{lem_propW1}, 
    $F$ is rationalized by a normalized welfare function $W:\mathbb{R}^n_+  \rightarrow \mathbb{R}$ such that $W$ is weakly monotone, symmetry, homogeneous and continuous and satisfies \eqref{eq_commonadd}.

    Let $M \subsetneq N$ be a nonempty set and $x, y \in \mathbb{R}^n_+$ such that $W(x_M, x_{N\backslash M}) > W(y_M, x_{N \backslash M }) $. 
    % Consider the two problems $\text{scmp} \{ (x_M, x_{N\backslash M}) , (y_M, x_{N \backslash M }) \}$ and $\text{scmp} \{ (x_M, y_{N\backslash M}) , (y_M, y_{N \backslash M }) \}$. 
    By Theorem \ref{thm_rational}, 
    \begin{equation*}
        (x_M, x_{N\backslash M}) \in F( \text{scmp} \{ (x_M, x_{N\backslash M}) , (y_M, x_{N \backslash M }) \} )
    \end{equation*}
     and 
     \begin{equation*}
         (y_M, x_{N\backslash M}) \notin F(\text{scmp} \{ (x_M, x_{N\backslash M}) , (y_M, x_{N \backslash M }) \}). 
     \end{equation*}
    By \textit{separability for EAI}, $(y_M, y_{N\backslash M}) \notin  F(\text{scmp} \{ (x_M, y_{N\backslash M}) , (y_M, y_{N \backslash M }) \})$. 
    By \textit{intermediate Pareto} and \textit{anonymity}, $(x_M, y_{N\backslash M}) \in  F(\text{scmp} \{ (x_M, y_{N\backslash M}) , (y_M, y_{N \backslash M }) \})$. 
    Therefore, $W(x_M, y_{N\backslash M}) > W(y_M, y_{N\backslash M})$.

    Let $\widetilde{W}: \mathbb{R}^n \rightarrow \mathbb{R}$ be the unique extension of $W$ satisfying \eqref{eq_commonadd2} (Lemma \ref{lem_propW1}). 
    By the result of the last paragraph, for all $x, y \in \mathbb{R}^n$, 
    \begin{equation}
    \label{eq_semisepa}
         \widetilde{W} (x_M, x_{N\backslash M}) > \widetilde{W}(y_M, x_{N \backslash M }) \implies \widetilde{W}(x_M, y_{N\backslash M}) > \widetilde{W}(y_M, y_{N\backslash M}). 
    \end{equation}
   
    Next we verify that for all  nonempty subset $M \subsetneq N$ and $x, y \in \mathbb{R}^n$, $\widetilde{W} (x_M, x_{N\backslash M}) \geq  \widetilde{W}(y_M, x_{N \backslash M }) $ implies $\widetilde{W}(x_M, y_{N\backslash M}) \geq \widetilde{W}(y_M, y_{N\backslash M})$.
    If $\widetilde{W}(y_M, y_{N\backslash M}) > \widetilde{W}(x_M, y_{N\backslash M})$, then by \eqref{eq_semisepa}, $\widetilde{W}(y_M, x_{N\backslash M}) > \widetilde{W}(x_M, x_{N\backslash M})$. This is a contradiction. 

    Then, by applying Theorem 13 in \citet{blackorby2002utilitarianism}, $\widetilde{W} (x) = \sum_{i\in N } x_i$ for all $x\in \mathbb{R}^n$.\footnote{Note that  \citet{blackorby2002utilitarianism} imposed axioms on social welfare orderings defined over $\mathbb{R}^n$. Although we consider properties of the function $\widetilde{W}$, it is easy to rewrite them to the counterpart properties of social welfare orderings. 
    For the original result, see also \citet{maskin1978theorem}. } 
\end{proof}

%%%%%%%%%%%%%%%%%%%%%%%%%%%%%%%%%%%%%%%%%%%%%%%%%%%
\subsection{Proof of Proposition \ref{prop_originalKS}}
%%%%%%%%%%%%%%%%%%%%%%%%%%%%%%%%%%%%%%%%%%%%%%%%%%%

The first part can be proved by a minor modification of the proof of Theorem 3 in \citet{xu2006alternative}, and we can prove the second part straightforwardly. 
% Since we will refer to the first part of this proposition in the proof of Theorem \ref{thm_joint}, we provide a proof of it. 

\begin{proof}[\bf Proof of Proposition \ref{prop_originalKS}]
    Since it is straightforward to prove that the KS choice rule satisfies the axioms in the statements,  we only prove the only-if part of the first part. 
    Let $F$ be a choice rule that satisfies \textit{weak Pareto}, \textit{scale invariance}, \textit{strong symmetry}, and \textit{contraction independence for EAI}. 

   Take an arbitrary problem $X \in \mathcal{P}$. 
    By \textit{scale invariance}, we can assume $b_i (X) = b_j (X)$ for all $i,j\in N$. 
    Let $x \in \mathbb{R}^n_{+}\backslash \{ \mathbf{0}\}$ be such that $\{x_{(1)} \mathbf{1} \}$ is the set of outcomes chosen by the KS choice rule in $X$ and $X \subset \text{scmp} \{ x\}$. 
    By \textit{strong symmetry} and \textit{weak Pareto}, $\{ x_{(1)}  \mathbf{1} \} = F(\text{scmp} \{ x\})$. 
    Since $x_{(1)}  \mathbf{1} \in X \cap F( \text{scmp} \{ x\}) $, \textit{contraction independence for EAI} implies that $F(X) = X \cap F( \text{scmp} \{ x\})  = \{ x_{(1)} \mathbf{1}  \}$, as desired.  
\end{proof}

%%%%%%%%%%%%%%%%%%%%%%%%%%%%%%%%%%%%%%%%%%%%%%%%%%%
\subsection{Proof of Theorem \ref{thm_joint}}
%%%%%%%%%%%%%%%%%%%%%%%%%%%%%%%%%%%%%%%%%%%%%%%%%%%

In this section, we only consider the case where $n = 2$. 
We provide  a proof of Theorem \ref{thm_joint} and then explain why this proof cannot be applied in the $n$-person case where $n\geq 3$.

Let $F$ be a choice rule that satisfies \textit{weak Pareto}, \textit{scale invariance}, \textit{anonymity}, \textit{contraction independence for EAI}, \textit{continuity},  \textit{equal addition independence for EAI}, and \textit{compromisability for EAI}. 
For all $x, y \in \mathbb{R}^2$, let $l (x, y)$ be the line segment joining $x$ and $y$. 

\begin{lem}
\label{lem_joint1}
    Let $F$ be a choice rule that satisfies the axioms in Theorem \ref{thm_joint} but does not satisfies \textit{intermediate Pareto}. 
    Then there exists $x\in \mathbb{R}^2_+ \backslash \{ \mathbf{0} \}$ such that $x_1 \neq x_2$ and $\{ x_{(1)} \mathbf{1} \} = F(\text{scmp} \{ x \})$. 
\end{lem}

\begin{proof}
    Suppose that  $x\in F(\text{scmp} \{ x \})$ for all $x\in \mathbb{R}^2_+ \backslash \{ \mathbf{0} \}$. 
    Then \textit{intermediate Pareto} can be replaced by \textit{weak Pareto} in the proof of Theorem \ref{thm_rationalordering}-\ref{thm_main}, so $F$ is a relative fair choice rule. 
    However, it satisfies \textit{intermediate Pareto}, which is a contradiction. 

    Therefore, we can assume that there exists $x\in \mathbb{R}^2_+ \backslash \{ \mathbf{0} \}$ such that $x\notin F(\text{scmp} \{ x \})$. It is sufficient to prove that $x_1 \neq x_2$ and $\{ x_{(1)} \mathbf{1} \} = F(\text{scmp} \{ x \})$. 

    First, suppose to the contrary that $x_1 = x_2$.  Since $x\notin F(\text{scmp} \{ x \})$, \textit{weak Pareto} and \textit{anonymity} implies there exists $y\in \mathbb{R}^2_+ $ such that $y_2 = x_2 > y_1$ and $y\in F(\text{scmp} \{ x \})$. 
    Since $F(\text{scmp} \{ x \})$ is closed and $x\notin F(\text{scmp} \{ x \})$, we can take $y(\neq x)$  as maximizing $y_1$ in $F(\text{scmp} \{ x \}) \cap l(  x, (0, x_2) )$.\footnote{The fact that $F(X)$ is closed for all $X\in\mathcal{P}$ follows from \textit{continuity}. We briefly verify this. Let $X\in\mathcal{P}$. Define $\{ X^k \}_{k\in \mathbb{N}}\subset \mathcal{P}$ as $X^k = X$ for all $k\in N$ and $\{s^k \}_{k\in \mathbb{N}} \subset F(X)$ as an arbitrary sequence that converges to $x \in X$. 
    By \textit{continuity}, we have $x\in F(X)$. } 
    (See Figure \ref{fig_proofThm3-1}.)
    By \textit{anonymity}, $(y_2, y_1)\in F(\text{scmp} \{ x \})$. 
    By \textit{compromisability for EAI} and \textit{weak Pareto}, there exists $z\in  F(\text{scmp} \{ x \})$ and $z\geq  ((y_1 + y_2 ) /2 ,(y_1 + y_2 ) /2 )$ and $z\in  l( x, (x_1, 0) ) \cup l(  x, (0, x_2) )$. 
    By \textit{anonymity}, we can assume $z\in l(  x, (0, x_2) )$. 
    We have $z\in F(\text{scmp} \{ x \})$ but  $z_1 \geq (y_1 + y_2 ) /2 > y_1$.   
    This is a contradiction to the definition of $y$. Thus, we have $x_1 \neq x_2$. 

    \begin{figure}
        \centering
        \includegraphics[width=0.8\linewidth]{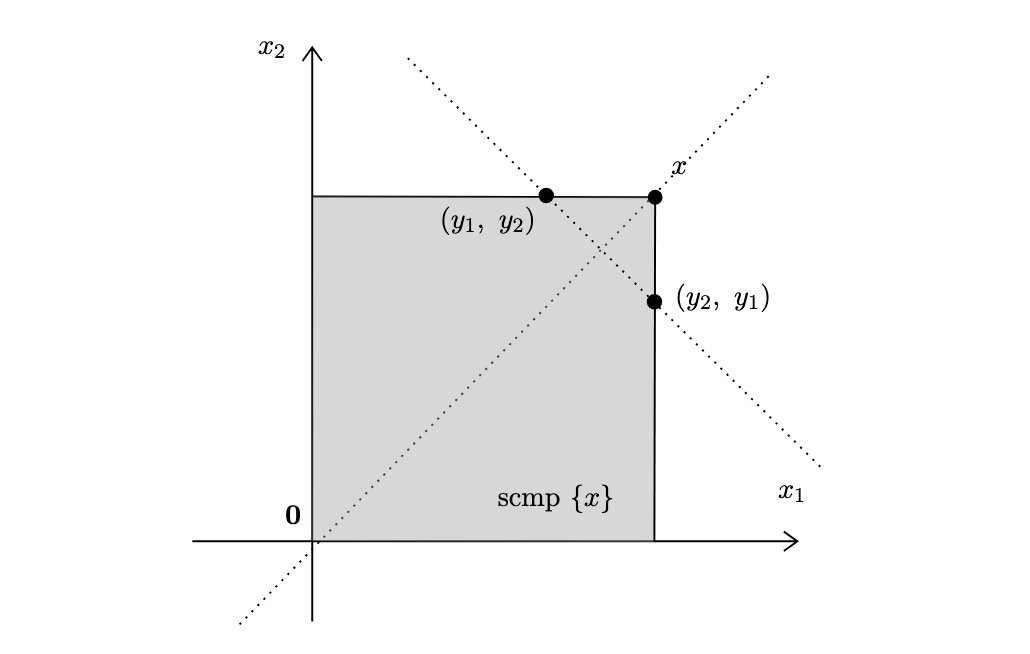}
        \caption{$\text{scmp} \{x \}$ when $x_1 = x_2$}
        \label{fig_proofThm3-1}
    \end{figure}

    We can assume $x_1< x_2$ without loss of generality. To prove $\{ x_{(1)} \mathbf{1} \} = F(\text{scmp} \{ x \})$, suppose to the contrary that there exists $y\notin \{ x, x_{(1)} \mathbf{1} \}$ such that  $y \in  F(\text{scmp} \{ x \})$. 
    By \textit{anonymity}, we can assume $y_1 < y_2$. 
    Let $(\alpha , \beta ) \in \mathbb{R}_{++} \times \mathbb{R}$ with $\alpha  y + \beta \mathbf{1} = x$. 
    By \textit{scale invariance}, 
    \begin{equation}
    \label{eq_y1_1}
        \alpha y \in F(\text{scmp} \{ \alpha x \}). 
    \end{equation}
    By \textit{equal addition independence for EAI}, 
    \begin{equation}
    \label{eq_y1_2}
         \alpha  y + \beta \mathbf{1} \in F(\text{scmp} \{ \alpha  x +  \beta \mathbf{1} \})
         \iff \alpha  y \in F(\text{scmp} \{ \alpha  x \}). 
    \end{equation}
    (Note that in the cases where $\beta< 0$, the above discussion holds.) 
    By \eqref{eq_y1_1} and \eqref{eq_y1_2},  $x = \alpha  y + \beta \mathbf{1} \in F(\text{scmp} \{ \alpha  x +  \beta \mathbf{1} \})$. 
    Since $x > y$  implies $\alpha  x +  \beta \mathbf{1} \geq \alpha y +  \beta \mathbf{1} =  x$, we have $\text{scmp} \{ x \} \subset \text{scmp} \{ \alpha  x +  \beta \mathbf{1} \}$. 
    By $x\in \text{scmp} \{ x \} \cap F(\text{scmp} \{ \alpha x +  \beta \mathbf{1} \})$ and \textit{contraction independence for EAI}, we have $x\in F(\text{scmp} \{ x \})$, which is a contradiction. 
\end{proof}

\begin{lem}
\label{lem_joint2}
    Let $F$ be a choice rule that satisfies \textit{weak Pareto}, \textit{scale invariance}, \textit{anonymity}, \textit{contraction independence for EAI}, \textit{continuity},  \textit{equal addition independence for EAI}, and \textit{compromisability for EAI}.
    If $\{ x_{(1)} \mathbf{1} \} = F(\text{scmp} \{x\})$ for some $x\in \mathbb{R}^2_{+}\backslash \{ \mathbf{0}\}$ with $x_1 \neq x_2$, then $\{ y_{(1)} \mathbf{1} \} = F(\text{scmp} \{y\})$ for all $y\in \mathbb{R}^2_{+}\backslash \{ \mathbf{0}\}$. 
\end{lem}

\begin{proof}
    Suppose that $\{ x_{(1)} \mathbf{1} \} = F(\text{scmp} \{x\})$ for some $x\in \mathbb{R}^2_{+}\backslash \{ \mathbf{0}\}$ with $x_1 \neq x_2$. 
    Let $y\in \mathbb{R}^2_{+}\backslash \{ \mathbf{0}\}$. 
    Since $\text{scmp} \{x\}$ and $\text{scmp} \{y\}$ are both symmetric problems, we can assume $x_1 < x_2$ and $y_1\leq y_2$. 
    Furthermore, by \textit{scale invariance}, we assume  $x_1 = y_1$ without loss of generality. 

    Consider the case where  $y_2 \leq x_2$. By \textit{contraction independence for EAI}, 
    \begin{equation*}
        \{ y_1 \mathbf{1} \} = \{ x_1 \mathbf{1} \} = F(\text{scmp} \{x\}) \cap \text{scmp}  \{y\} =  F(\text{scmp} \{y\}).
    \end{equation*}
    On the other hand, when $y_2 > x_2$, let 
    \begin{equation}
    \label{eq_defalpha}
        \alpha = {x_2 - x_1 \over y_2 - y_1} \in (0,1). 
    \end{equation}
    By \textit{scale invariance}, 
    \begin{equation*}
        F(\text{scmp} \{ y \}) = \qty( {1\over \alpha}, {1\over \alpha}  ) F(\alpha  \, \text{scmp} \{ y \}) = \qty( {1\over \alpha}, {1\over \alpha}  ) F( \text{scmp} \{ \alpha y \}).
    \end{equation*}
    Note that since $\alpha \in (0,1)$ and $x_1 = y_1$, we have $x_1 - \alpha y_1 > 0$. 
    By \textit{equal addition independence for EAI}, for all $z\in  \mathbb{R}^2_+$, 
    \begin{equation*}
        z \in F( \text{scmp} \{ \alpha y \}) \iff z + (x_1 - \alpha y_1) \mathbf{1}\in F( \text{scmp} \{ \alpha  y + (x_1 - \alpha y_1)  \mathbf{1} \}).
    \end{equation*}
    By \eqref{eq_defalpha}, we have $x = \alpha  y + (x_1 - \alpha y_1)\mathbf{1}$.
    Therefore,  for all $z\in \text{scmp} \{  y \}$, 
    \begin{align*}
        z\in F(\text{scmp} \{ y \}) &\iff \alpha z \in F( \text{scmp} \{ \alpha y \}) \\
        &\iff \alpha  z + (x_1 - \alpha y_1) \mathbf{1}\in F( \text{scmp} \{ x \}).
    \end{align*}
    Since $\{x_1 \mathbf{1}\} = F(\text{scmp} \{ x \})$, we have $\{y_1 \mathbf{1}\} = F(\text{scmp} \{ y \})$. 
    % On the other hand, when $y_2 > x_2$, let 
    % \begin{equation}
    % \label{eq_defalpha}
    %     \alpha = {y_2 - y_1 \over x_2 - x_1} > 1. 
    % \end{equation}
    % By \textit{scale invariance}, 
    % \begin{equation*}
    %     F(\text{scmp} \{ x \}) = \Bigg( {1\over \alpha}, {1\over \alpha}  \Bigg) F((\alpha \mathbf{1}) \, \text{scmp} \{ x \}) = \Bigg( {1\over \alpha}, {1\over \alpha}  \Bigg) F( \text{scmp} \{ (\alpha \mathbf{1}) x \}).
    % \end{equation*}
    % By \textit{equal addition independence for EAI}, for all $z\in  \mathbb{R}^2_+$, 
    % \begin{equation*}
    %     z \in F( \text{scmp} \{ (\alpha \mathbf{1}) x \}) \iff z + (y_1 - \alpha x_1) \mathbf{1}\in F( \text{scmp} \{ (\alpha \mathbf{1}) x + ( y_1 - \alpha x_1) \mathbf{1} \}).
    % \end{equation*}
    % By \eqref{eq_defalpha}, we have $y = (\alpha \mathbf{1}) x + ( y_1 - \alpha x_1) \mathbf{1}$.
    % Therefore,  for all $z\in \text{scmp} \{  x \}$, 
    % \begin{align*}
    %     z\in F(\text{scmp} \{ x \}) &\iff (\alpha \mathbf{1}) z \in F( \text{scmp} \{ (\alpha \mathbf{1}) x \}) \\
    %     &\iff (\alpha \mathbf{1}) z + (y_1 - \alpha x_1) \mathbf{1}\in F( \text{scmp} \{ y \})
    % \end{align*}
    % Since $\{x_1 \mathbf{1}\} = F(\text{scmp} \{ x \})$, we have $\{y_1 \mathbf{1}\} = F(\text{scmp} \{ y \})$. 
\end{proof}

\begin{proof}[\bf Proof of Theorem \ref{thm_joint}]
     If $F$ satisfies \textit{intermediate Pareto}, then by Theorem \ref{thm_main}, $F$ is a relative fair choice rule. 

    Consider the case where $F$ does not satisfy \textit{intermediate Pareto}. 
      Let $X \in \mathcal{P}$.  By \textit{scale invariance}, we can assume $b_i (X) = b_j (X)$ for all $i,j\in N$. 
      Let $x \in \mathbb{R}^n_{+}\backslash \{ \mathbf{0}\}$ be such that $\{x_{(1)} \mathbf{1}\}$ is the set of  outcomes chosen by the KS choice rule in $X$  and $X \subset \text{scmp} \{ x \}$. 
      By  Lemma \ref{lem_joint1} and \ref{lem_joint2}, $\{ x_{(1)}  \mathbf{1} \} = F(\text{scmp} \{ x\})$. 
      Since $x_{(1)}  \mathbf{1} \in X \cap F( \text{scmp} \{ x\}) $, \textit{contraction independence for EAI} implies that $F(X) = X \cap F( \text{scmp} \{ x\})  = \{ x_{(1)} \mathbf{1} \} $, as desired.  
\end{proof}

We briefly explain why this proof cannot be applied to the cases where $n\geq 3$. 
In the final paragraph in the proof of Lemma \ref{lem_joint1}, we have chosen two parameters $\alpha$ and $\beta$ satisfying $\alpha  y + \beta \mathbf{1} = x$. 
This can be done because $x, y \in \mathbb{R}^2$: 
If $n\geq 3$, we cannot always choose $\alpha$ and $ \beta$ satisfying this equation.

%%%%%%%%%%%%%%%%%%%%%%%

\section*{Acknowledgement}

This paper is based on Chapter 1 of the author's master's thesis submitted to Hitotsubashi University.
The author is grateful to Walter Bossert,  Yukihiko Funaki, Chiaki Hara, Youichiro Higashi, Noriaki Kiguchi, Nobuo Koida, Satoshi Nakada, Koichi Tadenuma, Norio Takeoka, Tsubasa Yamashita, Shohei Yanagita, Naoki Yoshihara,  and  the audience at the 21st Game Theory Workshop 2024 (Kyoto University) and the 2024 Autumn Meeting of Japanese Economic Association (Fukuoka University) for their helpful comments.

\bibliographystyle{econ}
\bibliography{reference}

%%%%%%%%%%%%%%%%%%%%%%%%%%%%%%%

\end{document}